\documentclass[runningheads]{llncs}
\usepackage{graphicx}
\usepackage{cite} 
\usepackage{algorithm}
\usepackage{algorithmic}
\usepackage{amsfonts,amssymb}
\usepackage{bm}
\usepackage{multirow}
\usepackage{amsmath}
\usepackage{float} 
\usepackage{subfig}
\usepackage{booktabs}
\usepackage[flushleft]{threeparttable}
\makeatletter 
\g@addto@macro\TPT@defaults{\footnotesize} 
\makeatother

\begin{document}
\title{Modeling User Behaviors in Machine Operation Tasks for Adaptive Guidance}

\author{Chen Long-fei\inst{} \and
Yuichi Nakamura\inst{} \and
Kazuaki Kondo\inst{}}

\institute{Academic Center for Computing and Media Studies,\\
Kyoto University, Japan\\
\email{chenlf@ccm.media.kyoto-u.ac.jp}\\
}

\maketitle 

\begin{abstract}
An adaptive guidance system that supports equipment operators requires a comprehensive model, which involves a variety of user behaviors that considers different skill and knowledge levels, as well as rapid-changing task situations.
In the present paper, we introduced a novel method for modeling operational tasks, aiming to integrate visual operation records provided by users with diverse experience levels and personal characteristics.
For this purpose, we investigated the relationships between user behavior patterns that could be visually observed and their skill levels under machine operation conditions.
We considered 144 samples of two sewing tasks performed by 12 operators using a head-mounted RGB-D camera and a static gaze tracker.
Behavioral features, such as the operator's gaze and head movements, hand interactions, and hotspots, were observed with significant behavioral trends resulting from continuous user skill improvement.
We used a two-step method to model the diversity of user behavior: prototype selection and experience integration based on skill ranking.
The experimental results showed that several features could serve as appropriate indices for user skill evaluation, as well as providing valuable clues for revealing personal behavioral characteristics.
The integration of user records with different skills and operational habits allowed developing a rich, inclusive task model that could be used flexibly to adapt to diverse user-specific needs.
\keywords{Egocentric vision \and Machine operation \and Human behavior analysis \and Skill improving process \and Adaptive guidance  \and RGB-D \and Gaze \and Hotspots.}
\end{abstract}

\section{Introduction}
In the domain of assembling or operational tasks, the novice or first-time users usually need a guide on how to deal with different equipment and devices.
To provide efficient assistance, smart guidance systems have been adopted and evaluated in previous studies \cite{cognitivelearning, Youdo, DimaAR, GlaciARsystem, NilsAR, motherboard, ARblock1, ARblock2}.
Implementing such systems can optimize task processes, improve outcomes, save physical energy, reduce mental workload, and provide economic benefits \cite{ARsurvey, ARmaintenance}.

One of the most important points of the assistant system is the adaptability for meeting user's complex task demands, as addressed in \cite{ARpilot}: 
\begin{quotation}{``\emph{The ability to flexibly aid users' needs to be emphasized to broaden the applicability of wearable computers. All the pilots using the wearable computer wanted to be able to customize the procedure to their own way. The system should not limit the methods of completing a task to one.}''}\end{quotation} 
and in \cite{ARmaintenance}:
\begin{quotation}{``\emph{Future AR applications should tailor AR content to address the needs of each individual. The trade-off between the positive and negative aspects of overlaid AR content is likely related to the experience-level of the individuals. }''}\end{quotation} 
Therefore, effective guidance system should (i) provide a diversity of guidance patterns that is compatible with a sufficient variety of possible users and (ii) support what is needed by situational awareness during the task execution process.
In other words, we expect that the guidance content includes a task model,  which can be used to track the workflow and guide users at each step; meanwhile, it can cover a variety of operation methods, as well as details and explanations which could satisfy the needs of novice users.
We also expect a timely context-aware ability. 
The guiding method should be associated with user's operational states \cite{ARsurvey2} for providing instructions at the right timing, e.g., when they feel difficult, hesitant, unconfident, or before an operational step, during the step or simply ignore the current step to display the next.

Being fully aware of a user's skill level is the key to design appropriate guiding approaches \cite{ARmaintenance}.
Novices may require step-by-step instructions and a detailed explanation on each task procedure. Accordingly, a guidance method typically employed by professionals may fail to assist new users.
Concerning intermediate level users, it may be reasonable to offer more advanced ways of instructions.
While professionals may seek to customize guidance according to their specific ways \cite{ARpilot} aiming to receive explanation on only a few key steps.
To assist a variety of operators considering different skill levels, it is necessary to define extensive patterns of operational methods, as well as specifying comprehensive behavior details.

Experiences from multi-skill-level users could serve as an excellent source for smart guidance systems on content learning, task modeling, and user behavior understanding.
However, an effective model cannot be directly derived by simply integrating all users' experiences,
since novice records may contain a large diversity of operation manners and many unnecessary actions and errors. 
If used as is, the model could be noisy. 

Concerning enhance the multiformity of the task model, our previous works \cite{chen2019model, chen2019intergrating} summarized the operational task using hand-machine interactions, and then manually selected prototype baselines to integrates experts' and beginners' operational experiences into an extensive task model.
The method provided a solution for dealing with behavior diversity. 
However, the integration method requires prior knowledge of the skill level of users.
Assuming that operation experiences that collected by crowdsourcing, etc., predetermining the identities of professionals and novices is still an problem itself.
Moreover, it needs to be considered that a single person may have varying experience levels corresponding to different learning stages, or different skills among operating steps concerning the same experience.
To mitigate this challenge, we propose an approach that implies automatically estimating skill levels related to particular operational records for task modeling.
In other words, the goal of this study is to automatically determine which operational experience exhibits high skill, and which experience is performed by a novice-like operator, as well as how the skills differ in different operational steps.

To perform this investigation, we systematically collect operating behavior patterns of multiple young adult participants who gradually learn and continuously improve their skills concerning two machine operational tasks.
We acquire operational records using a head-mounted RGB-D camera and a fixed gaze tracker.
The relationship between participant gaze, head, hands, and operation locations (hotspots) are automatically measured to analyze implicit user behavior patterns.
This multi-operator learning process can facilitate revealing the general characteristics of behavior in the process of skill development by considering interpersonal differences.
Our experiments successfully showed close relationships between operational behaviors, operational difficulties, and operator skill levels, such as task execution time and head motion can serve as strong indicators of user skills.
We also proposed an automatic experience selection method to build the prototype for task modeling, and focus on the interpretability of user behavior patterns.

\begin{figure}
\centering
\includegraphics[width=0.9\textwidth]{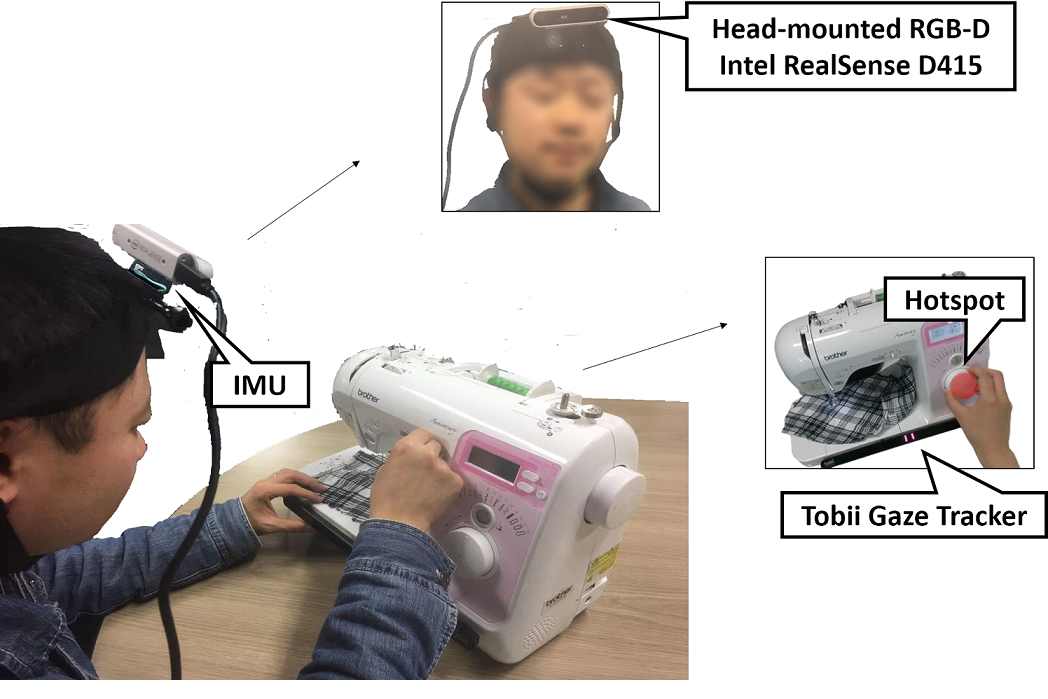}
\caption{A participant using a sewing machine. We analyze the operational behaviors of users by using their head, gaze and hand movements, and hotspots. Here, the goal is to find behavioral clues that indicate the skill level of the operator and the difficulty of the task.} \label{fig:framework}
\end{figure}

The machine operation environment is illustrated in Fig. \ref{fig:framework}.

\section{Related Work}
\noindent\textbf{Skill learning  }
Several studies \cite{fivestage, shuhari, six} described human skills acquisition as a complicated process with many learning stages, concerning both conceptual and behavioral aspects.
A traditional Japanese expression describes a 3-step learning process as ``Shu-Ha-Ri'' \cite{shuhari}, which can be translated as ``obey, break, and create.'' Obeying the rules and facts is suitable only for the early stages of learning.
A similar theory has been presented in \cite{fivestage}. In this study, skills acquisition is defined as five stages: novice, advanced beginner, competence, proficiency, and expertise.
The learning process of a novice was referred to as "\emph{being contingent on concept formation and the impact of fear, mistakes, and the need for validation}'' \cite{six}.
New joiners need to learn the facts and rules for determining the action.
As skills are getting improved, learners begin to apply personal judgment and emotions, aiming to adapt to various learning situations.
Upon becoming professionals, learners can disengage from the rules and emotions and directly identify the most efficient way to achieve the goal, and their behavior becomes organized and well-planned.
Unlike professionals' delicate behaviors, the novice learning process can provide essential behavioral clues for analyzing user skills.
Considering the guidance purpose, it is ideal for providing appropriate information to all users, but the highest priority is to support users from novices to intermediate levels.
Therefore, we aim to concentrate on the process of gradually improving skills from first-time users.

By investigating behavior patterns changing through the skill acquisition process, we can find critical behavioral characteristics that would help to evaluate their skills.
Several studies focused on investigating the relationships between operational behavior patterns and skills in a variety of applications.
In a research work on factory manufacturing \cite{methodEngineering}, it was observed that the task execution time monotonically decreased after several trials; however, this decreasing trend was not smooth, meaning that it could be segmented into several \emph{steep} and \emph{flat} stages.
Several studies outlined that the quality of various actions, such as accurate pose, economy, and fluidity of action movements, may indicate a high level of skill concerning daily operational tasks \cite{QQ} and also surgical environments \cite{relativehmm, handmotion}.
The study presented in \cite{OSATS} proposed several criteria for manual evaluation of operational quality between expert and novice surgeons, such as the performance time, the speed of using instruments, and the number of errors or procedure repetitions.
Khan et al. \cite{eyediff} considered gaze features to demonstrate the significant differences between novices and experts while watching videos of surgeries.
These studies determined useful features for assessing operating skills; however, there are still several open questions, such as which features can be considered sufficient for assessing various user skill levels.

\noindent\textbf{Operational behavior features  }
In the related studies, the researchers often consider user head movements, gazing, hand motions, objects, and their interrelationships as the critical cues to measure temporal and spatial behavior patterns concerning everyday operational tasks \cite{inwhatways, coordination, tessid, predictgaze}.
Land and Hayhoe \cite{inwhatways} investigated the correlation between the eye and hand movements concerning tea- and sandwich-making tasks.
Pelz et al.\cite{coordination} monitored the eye, hand, and head coordination during a block-copying task.
These studies demonstrated the temporal synergistic interrelation between eyes, head, and hand movements.
The distance between a gazing target and an operating location could also serve as an appropriate characteristic.
Skilled operators often shift their gaze to the next location before finishing the current ongoing process, for example, on average, 0.61 s in a tea-making task as measured in \cite{inwhatways}, and 100 ms earlier for a more skilled player compared with the less skilled one while playing cricket \cite{play}.
The presence of a strong correlation between gaze and head movements was demonstrated in several research works.
The studies \cite{tessid, predictgaze} proposed that head orientation could be considered a suitable approximation of gaze. They relied on strong coordination among eye, head, and hand movements in object manipulation tasks concerning egocentric gaze prediction.

Several studies adopted the concept of deep features to evaluate skills.
Doughty et al. \cite{doughty2019pros} proposed a supervised deep ranking model to determine skills in a pairwise manner. They noted that skill levels could differ depending on task procedures, and attention modules were added to focus on skill-relevant parts.
Parmar and Morris \cite{whatandhowwell} investigated spatiotemporal representations of motion and appearance using 3D convolutional neural networks (CNNs) to assess action quality in multiple diving tasks.
Li et al. \cite{spatialattention} adopted a spatial attention model based on a recurrent neural network (RNN) to assess hand manipulation skills.
In these research works, deep features were considered as complex characteristics corresponding to the spatiotemporal patterns of gaze, hand motions, objects, and their interrelationships; however, direct contribution of deep features was mostly difficult to analyze and apply to explain how human operational behavior differed depending on skill levels. Therefore, there are still open research problems, as outlined below.

\noindent\textbf{Open issue in skill assessment  }
Skill assessment and learning issues correspond to a combination of multiple factors.
First, required human actions and skills depend on the characteristics of a task. Skill evaluation methods may need to be modified for different tasks.
Second, the skill learning process is nonlinear and includes various complex stages.
Then, significant differences between the various operational steps of a task can be presented. Different individuals may have acceptable or poor performance in a same action. The skill level does not necessarily advance in a gradual manner across all steps, and accordingly, the type of required guidance is not uniform throughout the sequence of operations.
Lastly, operators with different personalities and physiques may have significant interpersonal differences in skill level progress.
Regarding the aforementioned observations, in the present study, we consider the primitive conventional features that are easy to measure, including gaze, head, hand and hotspots, which are common to many steps of various tasks.
It is evident that deep features can be applied efficiently when tuning for specific task; however, complex features are challenging to analyze and explain user behavior patterns.
We focus on task-independent features and consider using deep features for future investigation.

\section{Key Idea}
Our goal is to develop a comprehensive machine operating model from different users with different operating habits that can be used to guide them.
Understanding user skills enables an automated process of task modeling and guidance.
Detailed operation behaviors are greatly affected by user experience levels.
For example, operators who spend a lot of time searching for items and reviewing results tend to be less skilled. In contrast, operators who perform tasks efficiently with minimal steps are considered more skilled.

To provide a comprehensive understanding of behavioral differences among various skill levels and personal characteristics, we asked novice operators without any preceding experience to participate in the research and recorded their ongoing learning processes.
We aim to identify reliable features for skill assessment concerning spatio-temporal behavioral patterns of the learning process by acquiring the data in a fully automated way.

We examine quantitative and qualitative variations of those behavioral features, analyze how they change during learning, and differ among various persons.
A target of gaze is one of the most critical spots. It is a reliable cue of the operator's attention.
The hand is another vital spot, which indicates the operators' intentions and procedures to perform operations.
The third prominent spot is a hotspot, representing the frequent-interacting location between a hand and a machine.
Touching at the center of view may represent careful and concentrating operation, and touching far from the center may mean operation step with little attention.
Similarly, the duration, distance, and correlations of the user's gaze and hand moving towards hotspots are also used.
Strong correlations between behavioral features and skill improvement mean that they can provide clues about user skill levels.
However, it is necessary to consider interpersonal differences based on individual characteristics, learning speed, and physical dexterity.
If these features are not heavily dependent on individuals, we can reliably utilize them as skill level indices.
Conversely, personal characteristics can be distinguished according to the features that are diverse among users and are less correlated with the learning progress itself.

In addition to the above analysis of multiple users, we asked users to rate each operation's difficulty in each experience. We investigated how the subjective perception of the task procedures' difficulty varies among users and how such perceptions change through learning.
It should be outlined that the proposed approach differs from previous research works that simply classify or rank skills among the recorded experience instances \cite{whosbetter, whatandhowwell, spatialattention}.
We suggest that extracted behavioral features can also enhance the interpretability and online learning ability of guidance systems.

\section{Operational Behavior Detection} \label{detection}

\subsection{Operation Environment and recording}

We select the tasks associated with operating a sewing machine as suitable representative operations using daily machines. 
Operations of a sewing machine cover various common action patterns, such as push, slide, rotate, seize, and cut. 
Some of these actions require practicing for new users.
The participants were asked to seat in front of a table and to operate a sewing machine having all required materials setup with reach of a hand. 
Unlike in the case of operating general objects, the machine surface may lack texture or shape.

We involved 12 participants, 6 men and 6 women at the age of 20's, and recorded operational trials concerning two different sewing tasks. As a result, a total of 144 trials were registered. 
Note that none of the participants had experience using a sewing machine before recording. 
Task 1 is ``sew a specific symbol'' that consists of 11 standard operation steps (procedures) in the official manual, and task 2 is ``cut the thread and restore the machine to initial state'' that includes 5 standards steps. 
Each participant performed the two tasks alternately and repeatedly.
A total 12 trials are recorded for each participant.

To capture the operational experience, we take the advantage of egocentric vision using a head-mounted RGB-D camera (Intel RealSense D415 \cite{D415}). 
A fixed gaze tracker (Tobii Eye Tracker 4C \cite{Tobii}) was set at the machine surface (Fig. \ref{fig:framework}) to continuously capture the user gaze during the task process at 30$fps$.

\emph{Operational unit (OU)  }
In object-related operations, eyes are often involved in identifying objects for future use and planning operations to be performed on such objects \cite{inwhatways}.
Based on this observation, we define a basic Operational Unit for each operation step as the sequence of three periods: "\emph{pure-gazing (saccade/fixation)}'', ``\emph{hand-approaching}'', and ``\emph{operating}".

The \emph{pure-gazing} period is the period between the end of the previous physical hand--machine contact and the moment the hand appears within sight range.
The \emph{hand-approaching} period is the period between the end of a pure-gazing period and the time at which the hand operation begins.
The \emph{operating period} is the period in which physical touches occur.

\begin{figure}
\centering
\includegraphics[width = 0.95\textwidth]{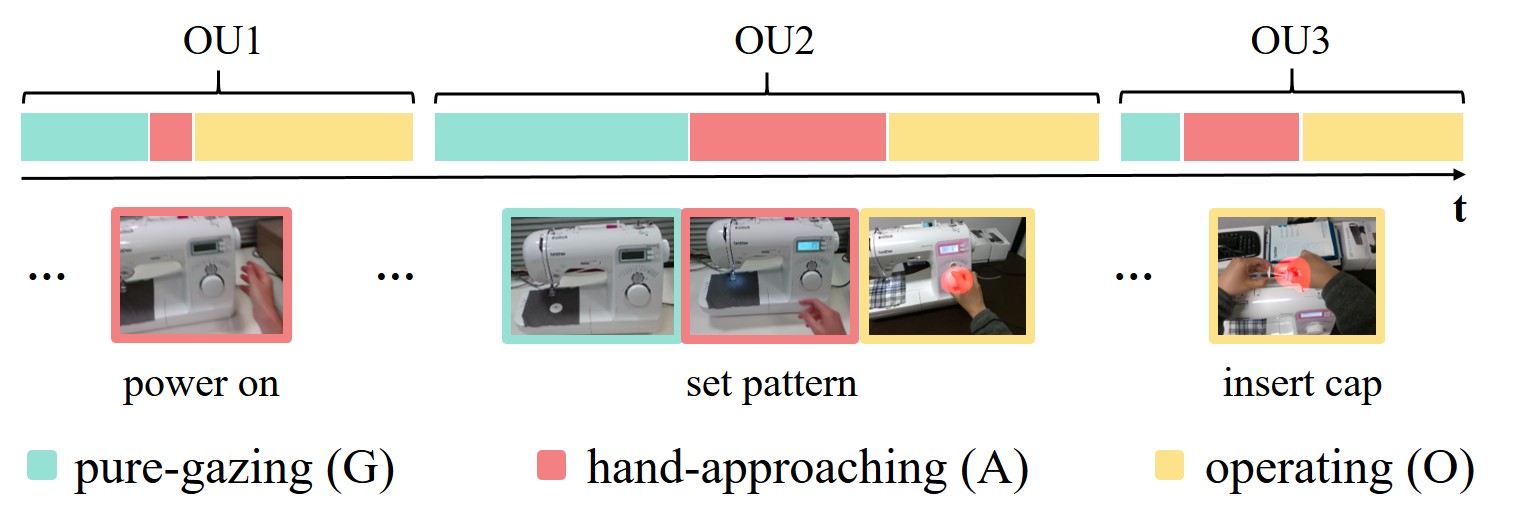}
\caption{The operational units corresponds to the operation steps.}
 \label{fig:OU}
\end{figure}

Each record of operational experience can be divided into a sequence of such OUs.
Detailed behaviors of each period may vary; for example, in some OUs, a gaze period is skipped, or two or more staff are gazed at sequentially. Our modeling of operational units deals with those variations as the duration is zero, and the duration is the sum of gazing periods for two or more staff, respectively.
An example is shown in Fig. \ref{fig:OU}.

\subsection{Visual Features}
Analyzing the egocentric vision and gaze tracking data, we were able to extract basic features corresponding to each OU.

\noindent\textbf{Global map}
A 2D global map of the sewing machine surface was prepared beforehand. 
Every egocentric view was aligned on the global map, we mapped them using SIFT features and applying homography transformation.
Then, global locations of the detected visual features were obtained.

\noindent\textbf{Gaze }
The user gaze was captured as a sequence of locations on the gaze tracker view field.
We need to register the view field of the fixed gaze tracker corresponding to the region on machine surface.
To do that, we first calibrated the gaze tracker for each operator using calibration points on a computer screen.
We then applied the screen region (namely, the view field of the gaze tracker) to the machine surface by registering the same geometry.
Examples of gaze distributions around hotspots on the global map are shown in Fig. \ref{fig:gazedistri}.

\noindent\textbf{Hand and hotspots }
To detect a hand, we first performed the segmentation of the foreground by considering the common operation distance ($20-100 cm$). Then, an HSV skin-color model was constructed for each user at the initial period of operation.
Hand locations were captured in every frame by filtering them through the skin-color model, depth and FPV size restriction.
Hotspots,  as crucial interaction areas on the machine surface, were detected automatically by clustering hand--machine touches in the spatiotemporal locations.
Detailed descriptions of the above processes can be found in the literature \cite{hotspots}.

\subsection{Behavioral Features}

\noindent\textbf{Duration ($T$), distance ($D$), velocity ($V$), frequency ($f$), and variance ($\delta$) }
For each OU mentioned above, the absolute duration of each period was measured as a behavioral feature. 
We considered that the relative distances among the hand, gaze target, and the hotspot were the essential features to characterize behavior patterns. 
Distances were calculated in pixels during each OU on the 2D global map. 
In addition, we calculated the distance changing speed, its variance and frequency.
defined as follows:

\begin{equation}
\begin{aligned}
V &= \Delta(d), \\
\delta^2 &= E[(d - \bar{d})^2]\\
f &= \mathbb{C}(d)/T.
\end{aligned}
\end{equation}
Here, $d$ is the distance between two regions; $\Delta$ is its difference; $\bar{d}$ is its mean; $\mathbb{C}$ is the number of sign changes corresponding to a distance in a period.
The frequency was derived by dividing the sign change number by the period duration.

\noindent\textbf{Head movement }
Head motion is represented by angular velocity in the horizontal direction and the vertical direction, which is estimated using the global motion vector of the egocentric RGB-d camera \cite{predictgaze} as follows:
\begin{equation}
V_{head_{x}} = \arctan(V_{global_{x}}/s_{img}*s_{sensor}/f),
\end{equation}
where $s_{img}$ and $s_{sensor}$ are the size of the image in pixels and the size of the camera CMOS sensor in $mm$, respectively, and $f$ is the focal length in $mm$. $V_{head_{x}}$ and $V_{global_{x}}$ represent the components for $x$ direction.
We also calculate the correlation between gaze and head movement to investigate their synergy.

\noindent\textbf{Trends and variations }
We extracted the aforementioned features from all trial records. Then, we compared the change of those feature values in the continuous skill improving experiences.

The trend of a feature associated with the skill improving process for a participant was obtained as the sum of differences in percentage among all his/her trials as follows:
\begin{equation}
\mathbb{D_u} = \sum_{n=2}^{N} (f^n_u - f^{n-1}_u)/\bar{f_u}.
\label{eq:trend}
\end{equation}
Here, $f^n_u$ is the feature value of a user at $n^{th}$ trial, and $\bar{f_u}$ is the mean feature value of trials corresponding to this user.
Then, the differences were averaged for all participants concerning two tasks to get the overall trend of a feature.

The inter- and intra- person standard deviation of a feature is calculated as:
\begin{equation}
\delta_{intra} = \sqrt{ \frac{1}{N}\sum_{n=1}^{N} (f_u^{n} - \bar{f_u})^2 }/\bar{f_u}
\quad \text{and} \quad
\delta_{inter} = \sqrt{\frac{1}{U}\sum_{u=1}^{U}(\bar{f_u} - \bar{f})^2}/\bar{f}.
\label{eq:inter}
\end{equation}
Here, $\bar{f}$ is the mean of $\bar{f_u}$ for all participants.

\noindent\textbf{Correlations }
The correlations among behavioral features, skill improvements, and subjective difficulty provide useful information for task modeling and guidance design, i.e., features with strong correlation can be good indexes for user skill levels and operational difficulty.
However, give a precise score to the skill level of a user at a particular trial is difficult.
For this purpose, we assume that each operator's skill level does not decrease through the experience accumulation process.
Thus, we rank the skill level as ordinal data aligned by task trial, e.g., the skill of a user in trial 1 is considerably no better than the skill in trial 3.
\begin{equation}
skill: T_{n} \geqslant T_{n-1} \ldots \geqslant T_{2}  \geqslant T_{1}.
\end{equation}
The operational difficulty is obtained by a subjective rating of task procedures using a six-point scale.
Each participant was asked to rate each operational step's difficulty from 0 to 5 (easiest to most difficult) after each trial.
The difficulty score may change in different trials. For example, in trial 1, a user could find rotating a dial complicated; however, the same procedure may feel easier in subsequent trials. The change in perceived difficulty may caused by skill level improvement.

The correlation coefficient of features to the ordinal scale of skill level was calculated using Spearman's rank correlation \cite{spearman}, and the correlation coefficient of features to operational difficulty scores was derived using the Pearson correlation \cite{pearson}.
The correlation between feature values and skill levels is considered for each trial whereas the correlation of difficulty is considered for each step of the operation, as follows:
\begin{equation}
R = Cov(\mathbf{f}, \mathbf{l})/\sigma_{\mathbf{f}}\sigma_{\mathbf{l}},
\label{eq:corre}
\end{equation}
where $\mathbf{f}$ is the vector of feature values (ranks), and $\mathbf{l}$ is the vector of skill levels or difficulties.

\section{Behavior Analysis Result}
{\textbf{Parameters} }
For each operation step, we first detected the operating period based on touches to a hotspot. Then, we detected the pre-operating periods using gaze and hand clues as mentioned above in Section \ref{detection}.
To filter out noise caused by detection errors, we ignored touches to hotspots for less than the threshold duration ($<$0.3$s$).
To simplify the notation in the following sections, the pure-gazing, hand-approaching, and operating periods in the OU are denoted G, A, and O, respectively.

\subsection{Behavior Changes through Skill Improvements}

\begin{figure}
\centering
\subfloat[]{
\makebox[\textwidth][c]{\includegraphics[width=1\textwidth]{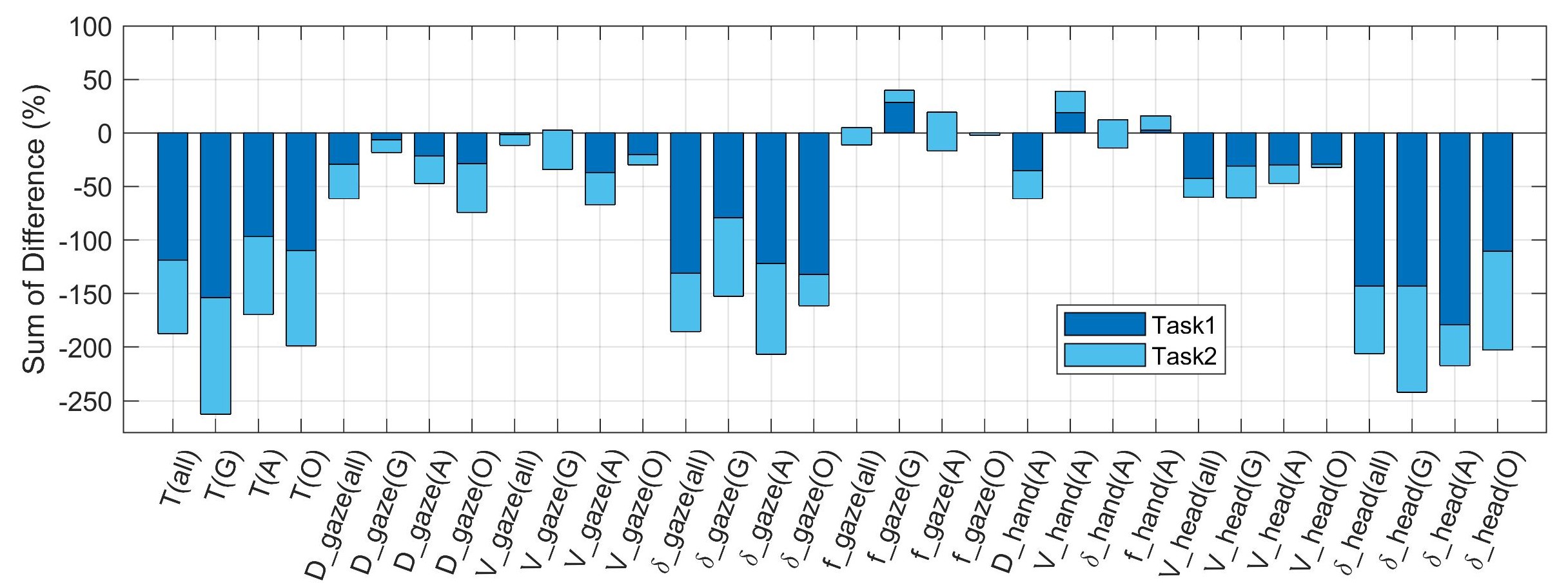}}}\\
\subfloat[]{
\makebox[\textwidth][c]{\includegraphics[width=1\textwidth]{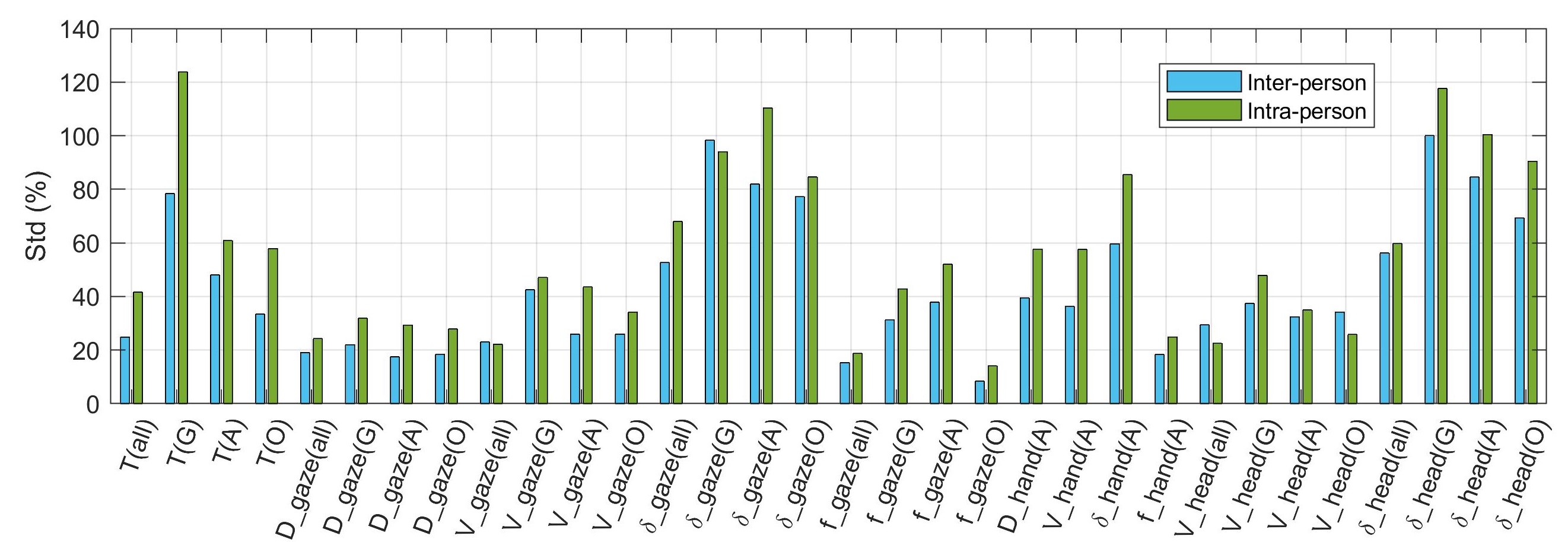}}}
\caption{(a) Overall trends of features (sum of differences) among trials and (b) inter- and intra-person standard deviations of features.}
 \label{fig:stddiff}
\end{figure}

\begin{itemize}
\item[$\bullet$]{\textbf{Overall trends and inter- and intra- person differences}}
\end{itemize}

We extracted the aforementioned features from all trial records. Then, we compared the difference between them in the continuous skill improving experience records. 

Fig. \ref{fig:stddiff} (a) represents the overall trends for the considered features by accumulating differences among trials averaged across the participants (calculated by Eq.\ref{eq:trend}).
Most of the features demonstrated same trends (negative or positive) in both tasks (28/32), suggesting that the changes of operational behavior features during the learning process were mostly task-independent in the conducted experiment.

Several features indicated an obvious downtrend as the learning proceeded, which are \emph{duration}, \emph{gaze variance}, and \emph{head variance}.
Specifically, from the first to the last trial, the most dramatically decreased features is \emph{duration of G} ($-262.6\%$), followed by \emph{head variance in G} ($-243.7\%$).
On the other hand, two features show an slight up-trend, which are \emph{hand velocity in A} ($+39\%$) and \emph{gaze frequency in G} ($+39.9\%$).

From an overall point of view, after skills are improved, the users could complete tasks faster, and the gaze and head movements became more stable. 
Moreover, the gaze target lied averagely closer to the operation locations, and hands approached the target faster.
The gaze frequency was slightly increased mainly because the significant reduction of duration.

The inter- and intra-person standard deviation of different features are shown in Fig. \ref{fig:stddiff} (b).
The standard deviation of individual features ranges from 8.46\% to 98.3\% for inter-person variations, and from 14.1\% to 123.8\% for intra-person variations, respectively.
Features show large variation among trials within a participant are similar to the their trend. 
Features show large variation among participants, that are, \emph{head variance} and \emph{gaze variance}, which can be used as good clues to indicate individual operation habits.

\begin{itemize}
\item[$\bullet$]{\textbf{Behavior differences in skill learning process (intra-person)}}
\end{itemize}

\begin{figure}
\centering
\makebox[\textwidth][c]{\includegraphics[width = 1.1\textwidth]{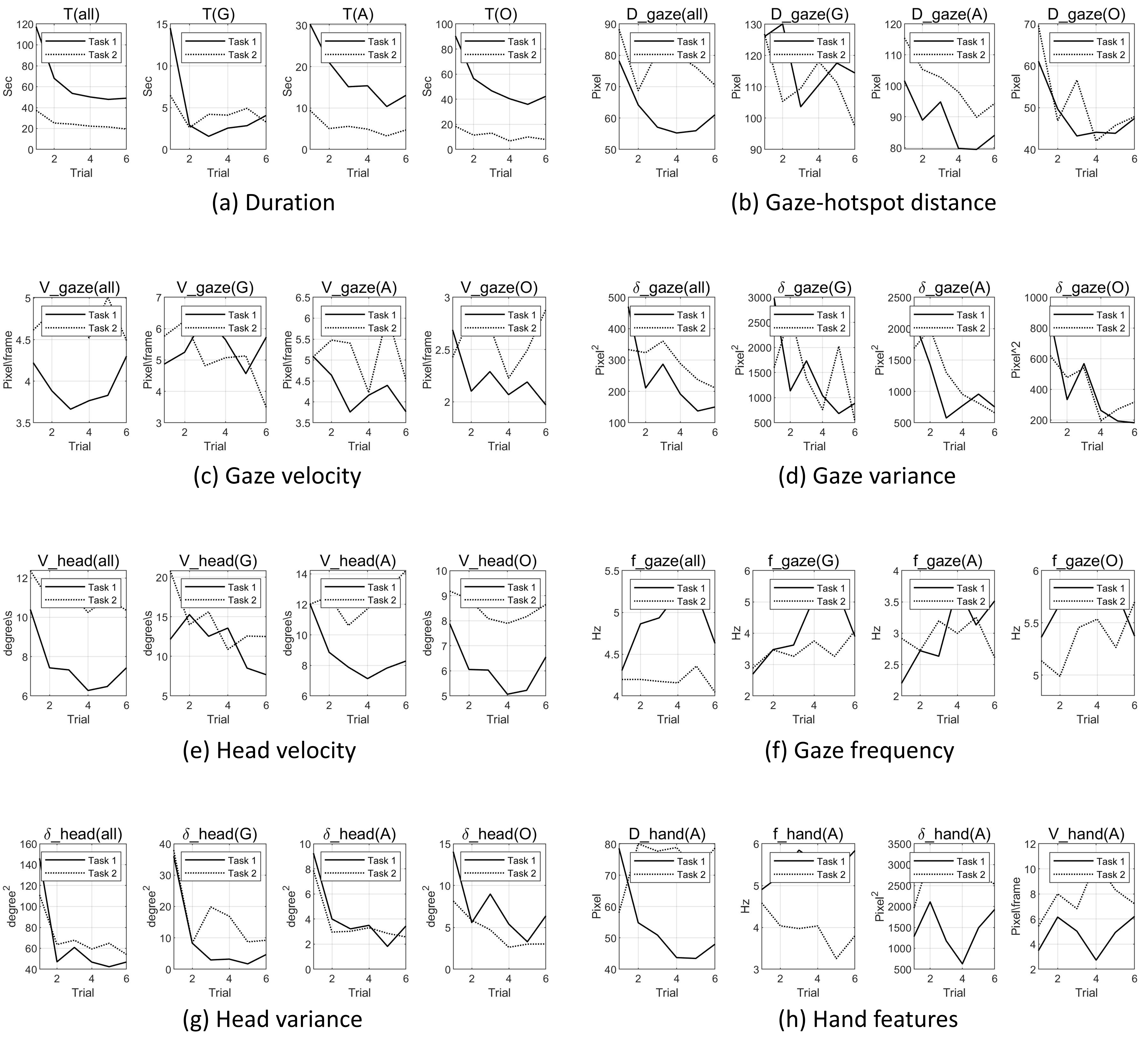}}
\caption{Detailed trends of each feature in two tasks.}
\label{fig:skilldetails}
\end{figure}

Fig. \ref{fig:skilldetails} shows the detailed trends of features from early to late trials.
Our main focus is on features where all participants show a significant trend as their skills improve, that are, {duration},  {gaze}, and {head}. 
While hand features did not show obvious trend with skill improving.

\noindent\textbf{{(1) Duration }}
The overall duration decreased monotonically for all participants in both tasks (Fig. \ref{fig:skilldetails} (a)).
For task 1, overall execution time of all participants decreased from an average of 120$s$ to 50$s$, and for task 2 overall execution time decreased from 40$s$ to 20$s$.
The largest percentage reduction was in G period; then in A period.
The duration decreased significantly within the first two trials; then, after two or more trials, the participants acquired most of knowledge underlying the considered operation tasks,  the time reduction was minor and smooth, reaching the minimum at the fifth trial.

The results show that low-skilled users required much more time to complete the task at initial trials. 
Especially before execution, they involved a lot of pure-gazing (search or hover) and a longer hand-approach time.
The speed up of decision making before starting execution and the quick reduction of physical operating time were continuously observed until the third trial.

Note that there demonstrates a slight upward trend of operation time at the sixth trials. 
Presumably, this is caused by the participants' intention to further improve of their performance.
One participant stated that he tried to stitch the symbol better by adjusting the cloth more carefully than in previous trials.

\noindent\textbf{{(2) Gaze }}
The average distance of gaze--hotspot is shown in Fig. \ref{fig:skilldetails} (b), and the overall gaze variance is shown in Fig. \ref{fig:skilldetails} (d).
The two features both overall decreased as user experience increased, especially the variance decreased dramatically for all OU periods.

At the initial trials, gaze had both large distance to operation location and large variance in all periods.
This was because novices required more checks to retain the relevant information prior to initiate an operation, and more extensive checks for result confirmation during the operation were required.
As skill improved, users located their gaze averagely closer to the operation location in both pre- and in-operation periods;
and the gaze movement range is much narrower.
On the other hand, the gaze frequency (f) and velocity (c) did not shown an monotonic trend.

We note that gaze--hotspot distance demonstrates a bowl shape for both tasks in the G period.
At the early trials, users did a lot of pre-operation search (large $D$ and $\delta$); 
in the middle trials, users tend to shift attention directly to the future target region (small $D$ and $\delta$); 
when getting more familiar in later trials, they did not need to concentrate on the specific target anymore (increasing $D$ while decreasing $\delta$).
This is because the prior knowledge for future operation positions was already formulated, and the users can rely on their memory to locate hotspots and guide their hands.

\begin{figure}[htbp]
\centering
\subfloat[]{
\makebox[\textwidth][c]{\includegraphics[width=0.8\textwidth]{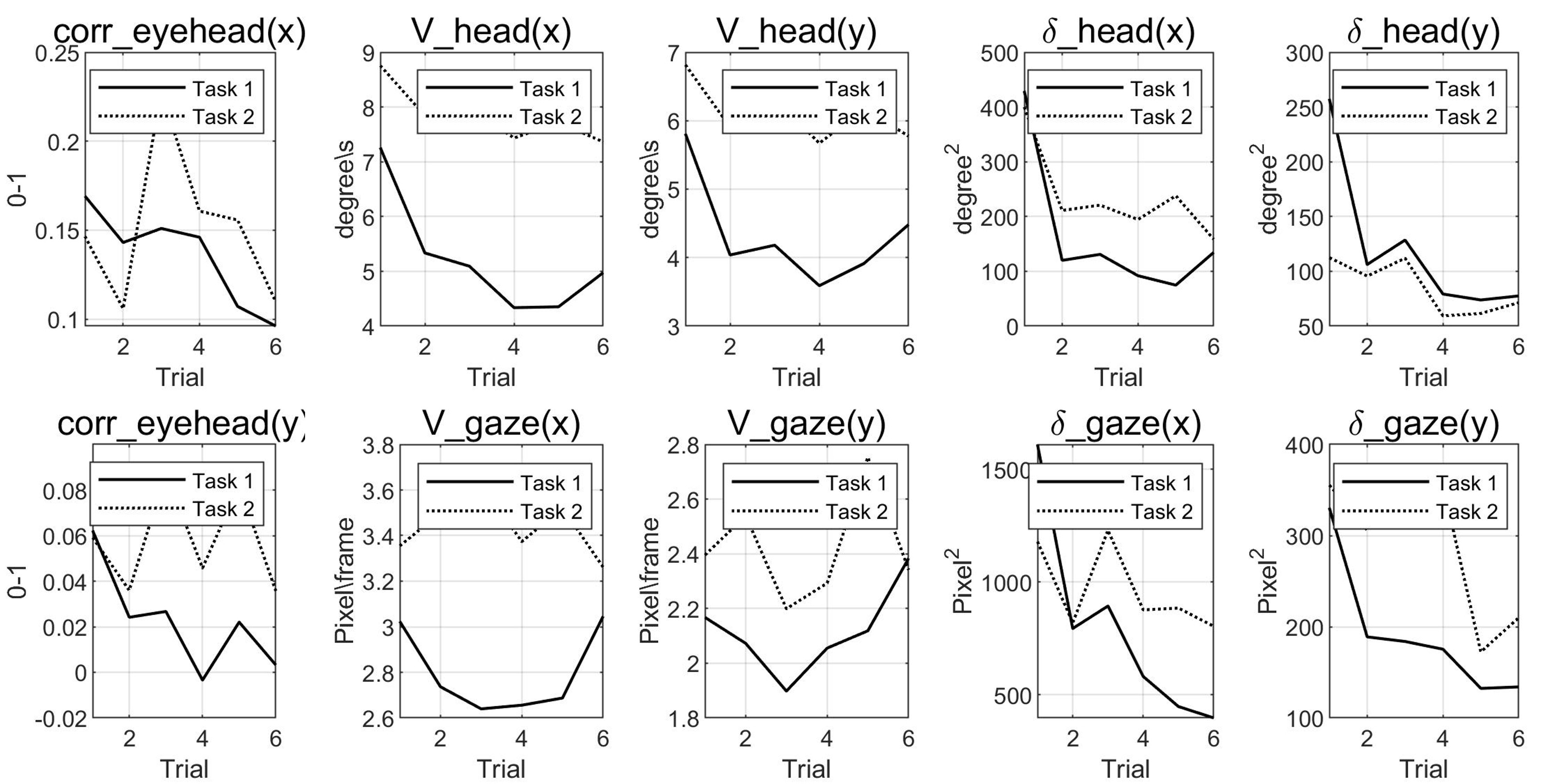}}}\\
\subfloat[]{
\makebox[\textwidth][c]{\includegraphics[width=0.8\textwidth]{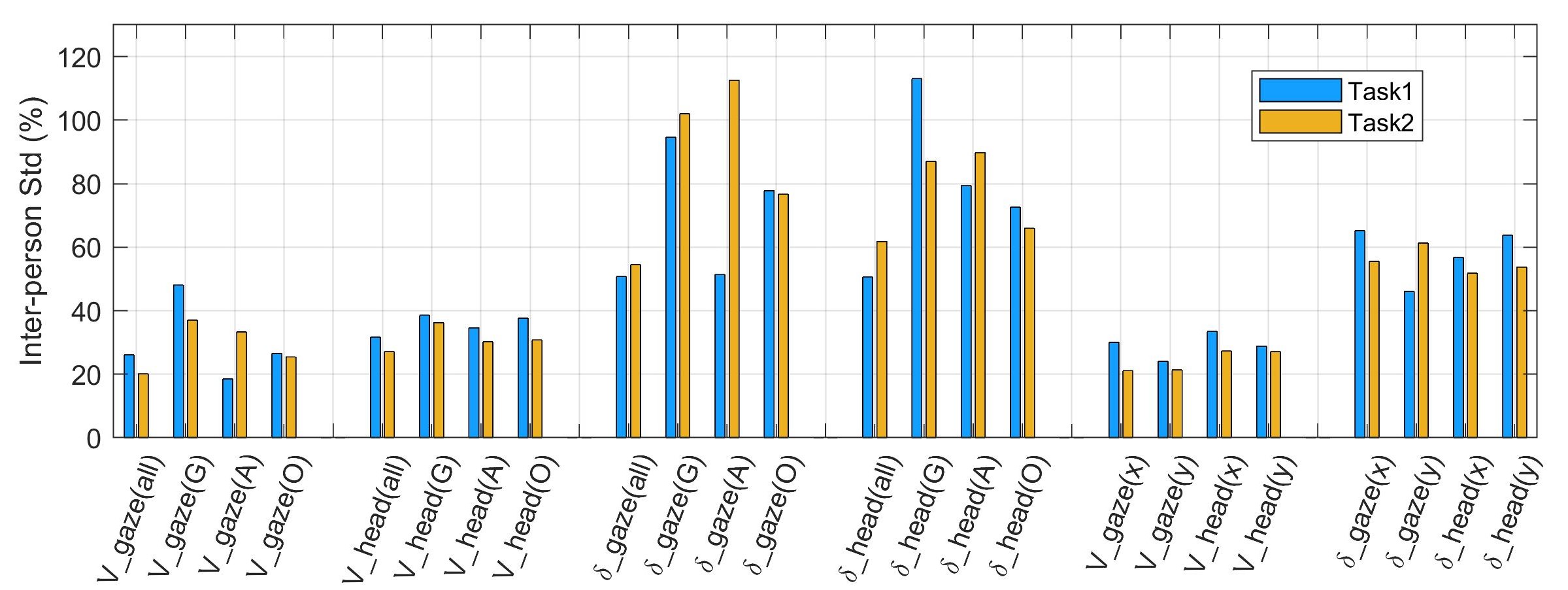}}}\\
\subfloat[]{
\makebox[\textwidth][c]{\includegraphics[width = 1\textwidth]{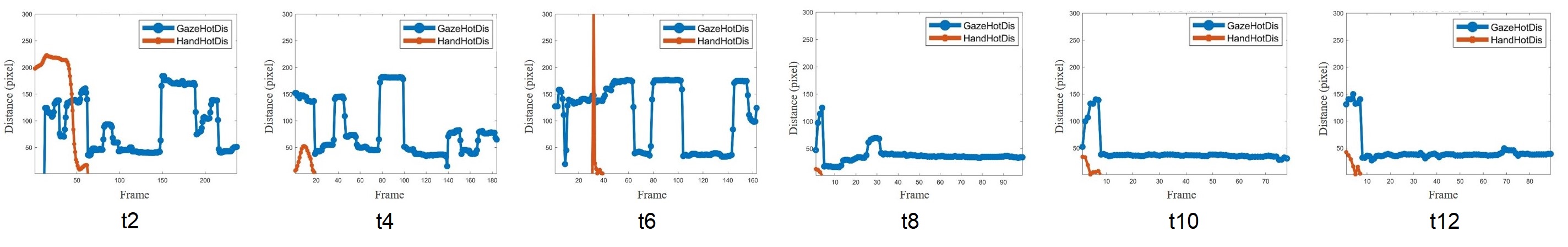}}}
\caption{(a) Comparison of the head and gaze features in horizontal ($x$) and vertical ($y$) directions. (b) Interpersonal standard deviations of the gaze and head features. (c) An example of changes in the gaze--hotspot distance and hand--hotspot distance of an operating procedure (\emph{rotate sewing pattern}) from a user, from trial 2 to trial 12.}
\label{fig:details}
\end{figure}

An example of changes of feature values from an user with skill improvement of the operating procedure \emph{rotate sewing pattern} is shown in Fig. \ref{fig:details} (c). We can note that this procedure's execution time is becoming shorter after more trials, while the gaze approaches the hotspot faster with less search during operation. 



\noindent\textbf{{(3) Head }}
The overall velocity and variance of head movement are both decreased monotonically from early to later trials for all users, as shown in Fig. \ref{fig:skilldetails} (e) and (g).
This shows that the stability, i.e., less motion, of the user's head could indicate a high skill level.
It show a similar trend with the duration, with all users drastically reducing their movements to low levels after the second trial. 
However, the reduction in subsequent trials is not as smooth as the duration, especially the head velocity in some middle trials.
This indicate the user's head movement is mainly reduced due to the reduction of pure-gazing behaviors in initial trials.

In conclusion, participants showed similar learning trends among trials. 
We can infer from first two trials that participants have acquired most of the knowledge needed to perform the task, and familiar with the actual operation through three trials.

%

\noindent\textbf{{(4) Gaze--head comparison }}
In particular, we compared the gaze and head features.
These two features had essential similarities of being capable of expressing user attention and intention, and it was possible to omit one or the other. As accurate gaze tracking usually required special devices \cite{tracker1, tracker2}, it would be advantageous in terms of cost if the head movements could be considered instead.

The head--gaze movement correlation scores estimated during the learning process are provided in Fig. \ref{fig:details} (a) column 1.
It could be seen that the gaze and head movements were almost uncorrelated in the vertical ($y$) direction and were weakly correlated in the horizontal one ($x$).
Considering the average score on the horizontal direction, the correlation almost monotonically decreased during the learning progress.
This is because the mental and physical costs of an eye movement is much less than moving the head.
Skilled users learned well the location of a target and tended to use the eye movement and to reduce the amount of head motion unless it was essential.
Therefore, as skills were improved, eye-head coordination become more asynchronous, which resulted in a difference between gaze and head behavior patterns.

Columns 2-5 Fig. \ref{fig:details} (a) represent the detailed trends in both horizontal and vertical directions corresponding to the head and gaze movements.
Head movement speed and variance were sharply decreasing in both directions.
However, the velocity of gaze movements did not represent an obvious trend concerning both tasks, while its variance decreased rapidly.

Fig. \ref{fig:details} (b) represents the interpersonal standard deviations of the gaze and head features corresponding to different OU periods (G, A, or O) and each direction ($x$ and $y$), as calculated by Eq.\ref{eq:inter}.
It could be observed that the variance demonstrated much larger interpersonal variation compared with the velocity of the both gaze and head features.
Specifically, the interpersonal variation of each direction was smaller than that of each OU period concerning both velocity and variance.
Therefore, the velocity of head and gaze from each direction ($x$ and $y$) could be considered as more reliable features to indicate skill levels among multiple users.

\begin{figure}[t]
\centering
\makebox[\textwidth][c]{\includegraphics[width=1\textwidth]{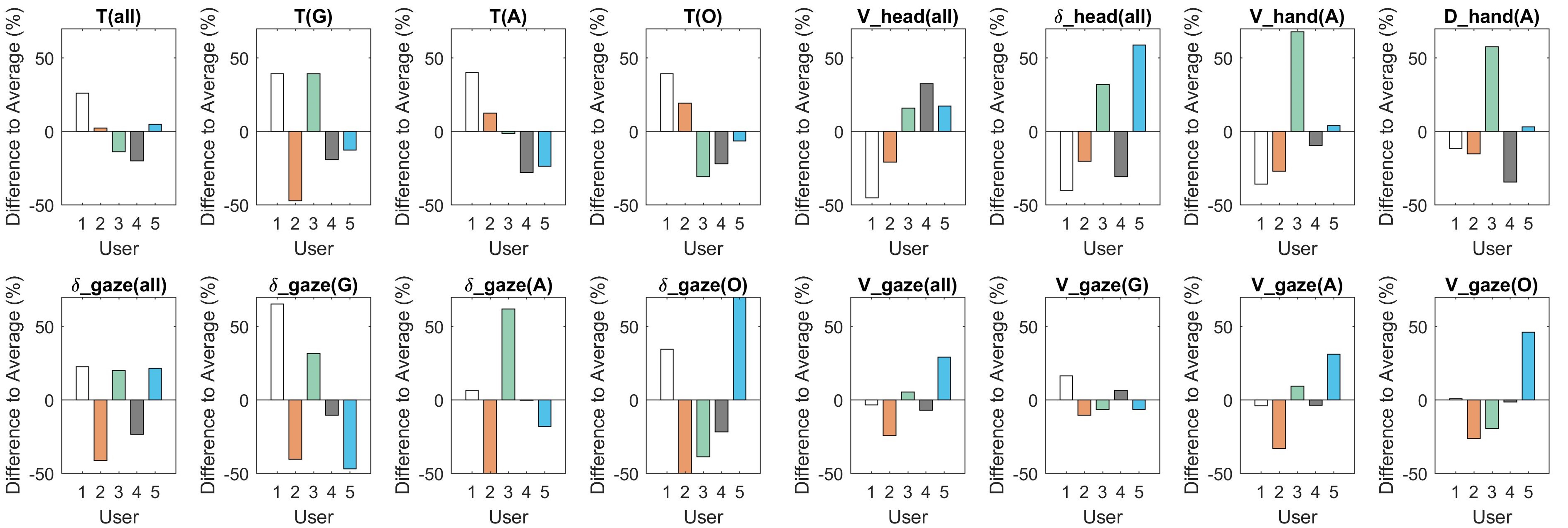}}
\caption{Personal behavior comparison of five participants in the different OU periods.}
\label{fig:personal}
\end{figure}

\begin{itemize}
\item[$\bullet$]{\textbf{Individual differences}}
\end{itemize}

Concerning the largest interpersonal variation, Fig. \ref{fig:personal} represents the detailed feature values for five users concerning \emph{head movement}, \emph{duration}, and \emph{gaze movement}. 

We noticed that Participant 1 had the longest operation time and large gaze variance in all periods, and the head and hand movements were also very slow. 
This indicated that this participant tended to search more or hesitated before taking action, meanwhile performing slowly during task execution. When providing guidance to such users, explicit instructions may be required.
In contrast, Participant 4 had the fastest head movement, the fastest task execution time, and a small gaze and head variance, which may indicate that the user was relatively skilled and might not require much guidance during the operation.

Participant 3 performed the considerable amount of search in the pre-operation periods (the longest G, and large gaze variance in G and A); however, he/she executed operations fast and concentrated on operation (gaze variance and velocity in O were small).
In contrast, participant 5 made decisions quickly prior to beginning the operation (small gaze variance in G and A), and often checked the progress or outcomes during the operation (gaze variance and velocity in O were large).
When providing guidance for these different types of users, timing was important, meaning that the guidance should have been provided at the appropriate moment during the process (for example, before beginning or during an operation).

Participant 2 had a short G period, while the gaze and head movements were very stable at all periods.
Such users with minor search behavior might not need any guidance before execution an operation.

We consider it difficult to only focus on a single feature to distinguish between the effects of interpersonal differences and learning stages for skill assessment.
From multiple features in fine-grained OU periods, we are able to observe and analyze the above interpersonal differences and identify various trends.


\begin{itemize}
\item[$\bullet$]{\textbf{Reliable clues to skill levels}}
\end{itemize}

We outline that the characteristics of a feature suitable for the evaluation of skill levels are:
a. Large difference between skill levels; b. monotone in terms of skill improvement; c. small interpersonal differences.

\begin{figure}[t]
\centering
\makebox[\textwidth][c]{\includegraphics[width=1\textwidth]{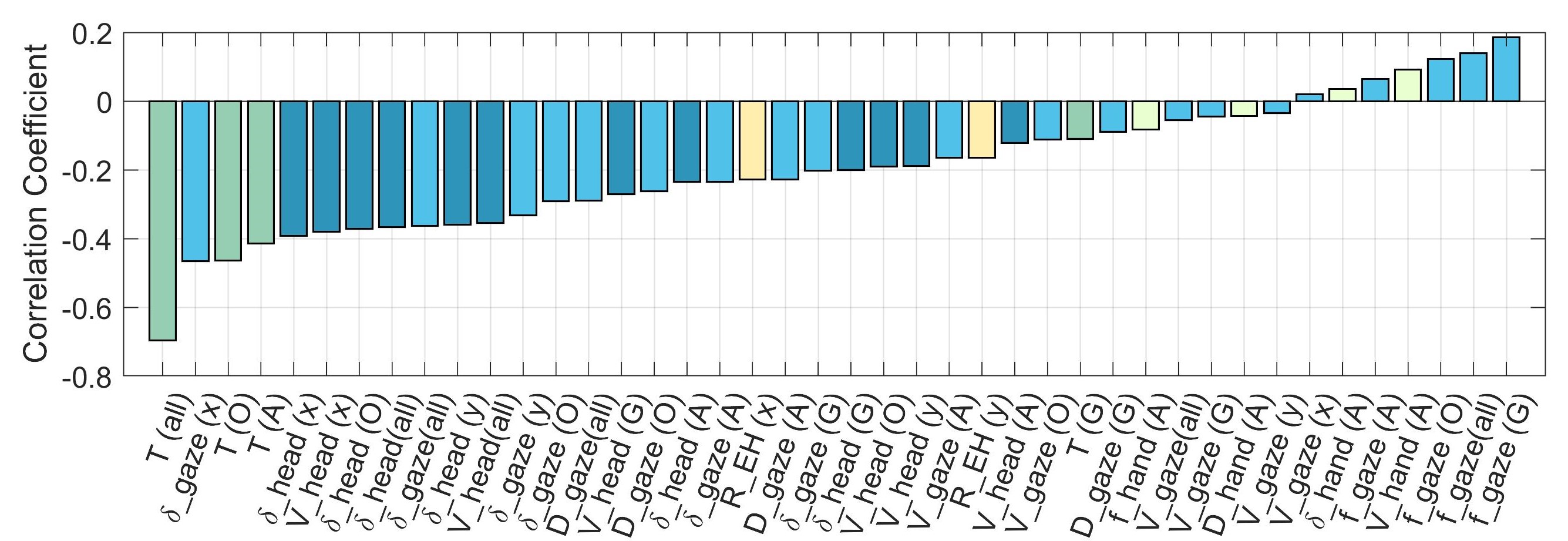}}
\caption{Correlation between all features and skills.}
\label{fig:correlationscore}
\end{figure}

The rank of correlation coefficients of all features corresponding to the skill levels (calculated by Eq.\ref{eq:corre}) are represented in Fig.\ref{fig:correlationscore}.
Among the features highly correlated to skills, the most relevant is \emph{duration (all)}, followed by \emph{gaze variance (x)},  \emph{head variance (x)}, and \emph{head velocity (x)}.
Compared with most related features, duration took 3 out of the top-5 features, and head features took 5 out of the top-10 features.
Concerning gaze, \emph{gaze--hostpos distance} was ranked behind \emph{gaze variance}.
We can see that the correlation of head features to skills was stronger than that of gaze features, and their interpersonal variation was relatively similar (Fig. \ref{fig:details} (b)). Therefore, considering the gaze features could be substituted by head features for the purpose of the present study.

\begin{figure}[t]
\centering
\makebox[\textwidth][c]{\includegraphics[width = 1\textwidth]{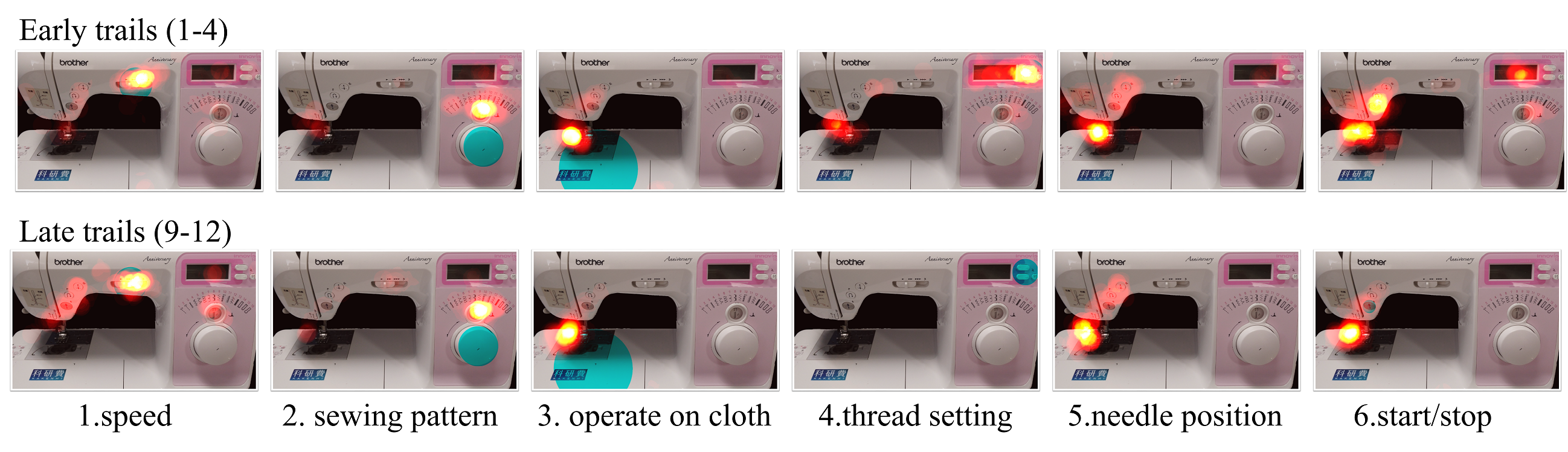}}
\caption{Gaze distribution difference (\emph{heat}) around hotspots (\emph{cyan}) in early and late trials for six different hotspots on the global map (accumulated among five participants).}
 \label{fig:gazedistri}
\end{figure}

\begin{itemize}
\item[$\bullet$]{\textbf{Gaze target differences}}
\end{itemize}
The user gaze location was distributed differently when operating the same hotspot during skill improvement, as shown in Fig. \ref{fig:gazedistri}.
The hotspot locations are denoted in \emph{cyan}, and the gaze locations are marked in \emph{red} and \emph{yellow} representing accumulated heat.
We compared the accumulated gaze location in earlier trials between those in later trials for five participants concerning 12 trials from both tasks. 

Regarding the overall gaze distribution, the user gaze was directed to a location where operational outcomes (effects) were occurring.
For example, when rotating the hotspot 2 (\emph{sewing pattern}), the gaze was directed to the pattern display panel above the dial. When operating on hotspots 3, 5, and 6 (\emph{cloth}, \emph{needle position}, and \emph{start$\backslash$stop}), the gaze was primarily directed towards the moving needle.
However, if there was no operational outcome region, the gaze was primarily located on the ongoing interacting region (hotspots 1 and 4).
This result supported the top-down control property of the gaze \cite{inwhatways}.

Then, we found that the user gaze located differently in different trials, which indicated that it could be suitable to evaluate user skill levels. 
Concerning hotspot 1 (\emph{speed}), we could see that, initially, the user gaze focused on an interacting region (trials 1--4), and after improving the skills (trials 9--12), the gaze shifted to the next operation region before the current operation was completed. 
We consider that this occurs because the interaction itself is rather simple without any operational outcome region. Once the users learned this step, they were capable of planning future steps and did not need to persistently concentrate on the current step. 
Concerning hotspot 6 (\emph{start$\backslash$stop}), it was observed that when the user skills were low (trials 1-4), they had to check both the button to interact with and the operational outcome region; however, in later trials (trials 9--12), there was no need to concentrate on the button.
Concerning hotspot 4 (\emph{thread setting}), in later trials, the users did not perform this operation because they learned that it was an unnecessary procedure.

Based on these observations, we can conclude that gaze of the low-skilled user tended to locate more on the current interacting regions, as they were not familiar with the current operational step. 
However, the operational outcome region always attracted considerable attention during an operation regardless of the user skill level.
In future research work, we consider focusing on the topic of quantifying the user gaze target to evaluate their skills and then, offering the suitable guidance content based on the gaze characteristics.

\begin{figure}
\centering
\subfloat[]{
\includegraphics[width=0.78\textwidth]{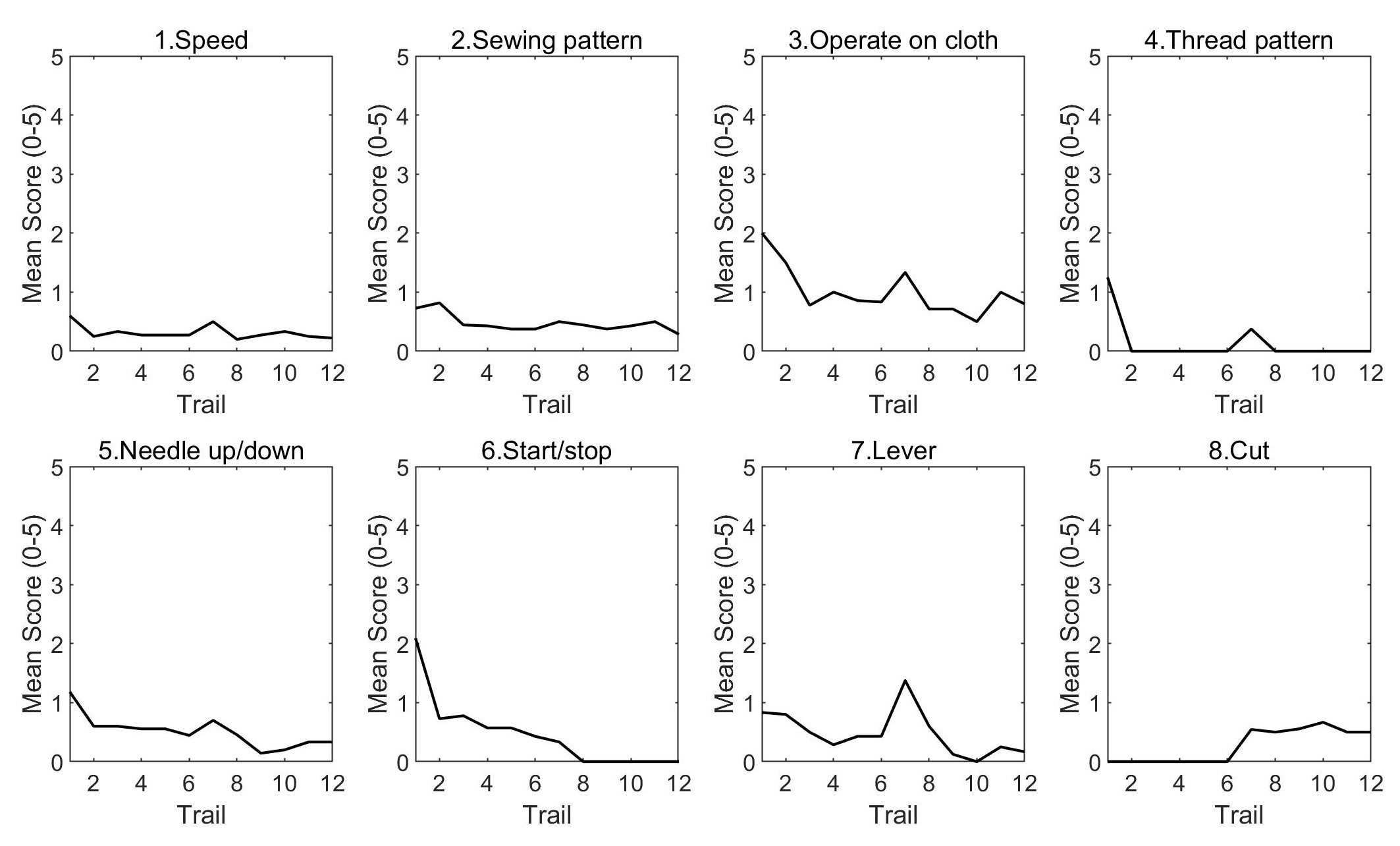}} \quad
\subfloat[]{
\makebox[\textwidth][c]{\includegraphics[width=0.8\textwidth]{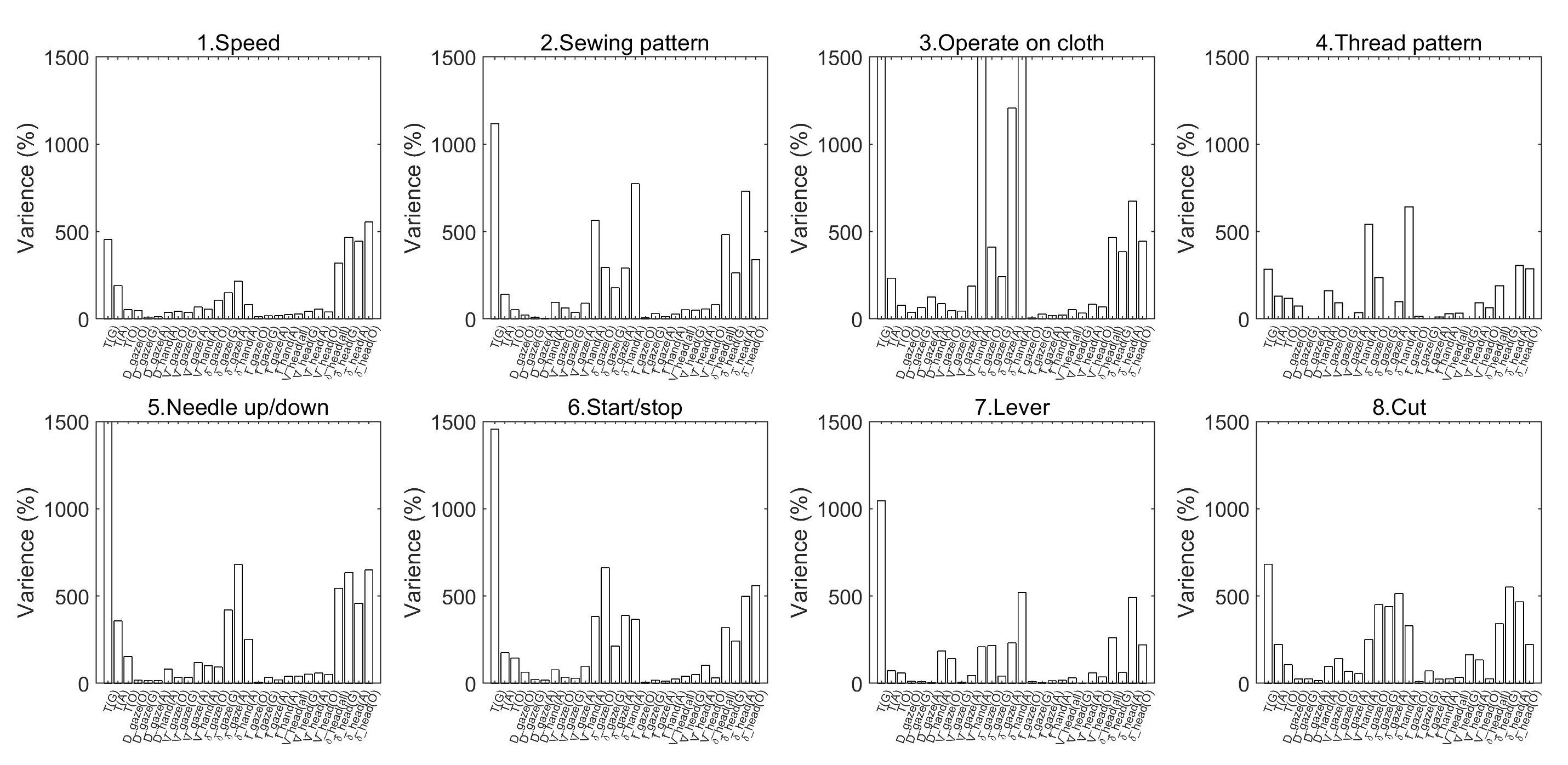}}}
\caption{(a) User ratings of operational difficulties for several steps with over increasing trials in both tasks, (b) feature variations in each hotspot (averaged by participant for all 12 trial of the two sewing tasks).}
\label{fig:difficultyrate}
\end{figure}

\subsection{Operational Difficulty}
After identifying reliable clues for skill levels, we examined differences among operation steps and their operational difficulties.
The difficulty of task steps could be a subsidiary hint for guidance offerings.

Fig. \ref{fig:difficultyrate} (a) shows the average score of user-rated difficulty in several main operation steps on 12 trials from both tasks.
The difficulties of some steps (e.g., steps 4 and 6) decreased sharply as the learning progressed.
For these kinds of operations, once a user knew how to perform them (e.g., \emph{push a button}), they were no longer considered difficult.
We refer to this type of difficulty as ``know-how difficulty.''
In contrast, some other operations are consistently rated difficult, e.g., steps 3 (\emph{sewing}).
We refer to this as ``skill-required difficulty.''
These types of operations may require a more comprehensive user guide, such as showing the details of a method or an alternative easier way.
In step 8 (\emph{cut}), a slight increase in difficulty was observed with skill improvement. 
Users find that thread cutting becomes relatively difficult (rated as 1) when most other operations are already easy (rated as 0).

Fig. \ref{fig:difficultyrate} (b) shows the distribution of feature values over all participants (depicted in terms of the variance of feature values).
The operations with relatively high difficulty scores (steps 3) demonstrate significant variance among features, especially for the \emph{duration of pure-gazing} and \emph{gaze/hand variance of approaching}.
This illustrates a bigger behavioral difference was observed among users when performing more complicated procedures.

\begin{figure}[htbp]
\centering
\includegraphics[width=0.8\textwidth]{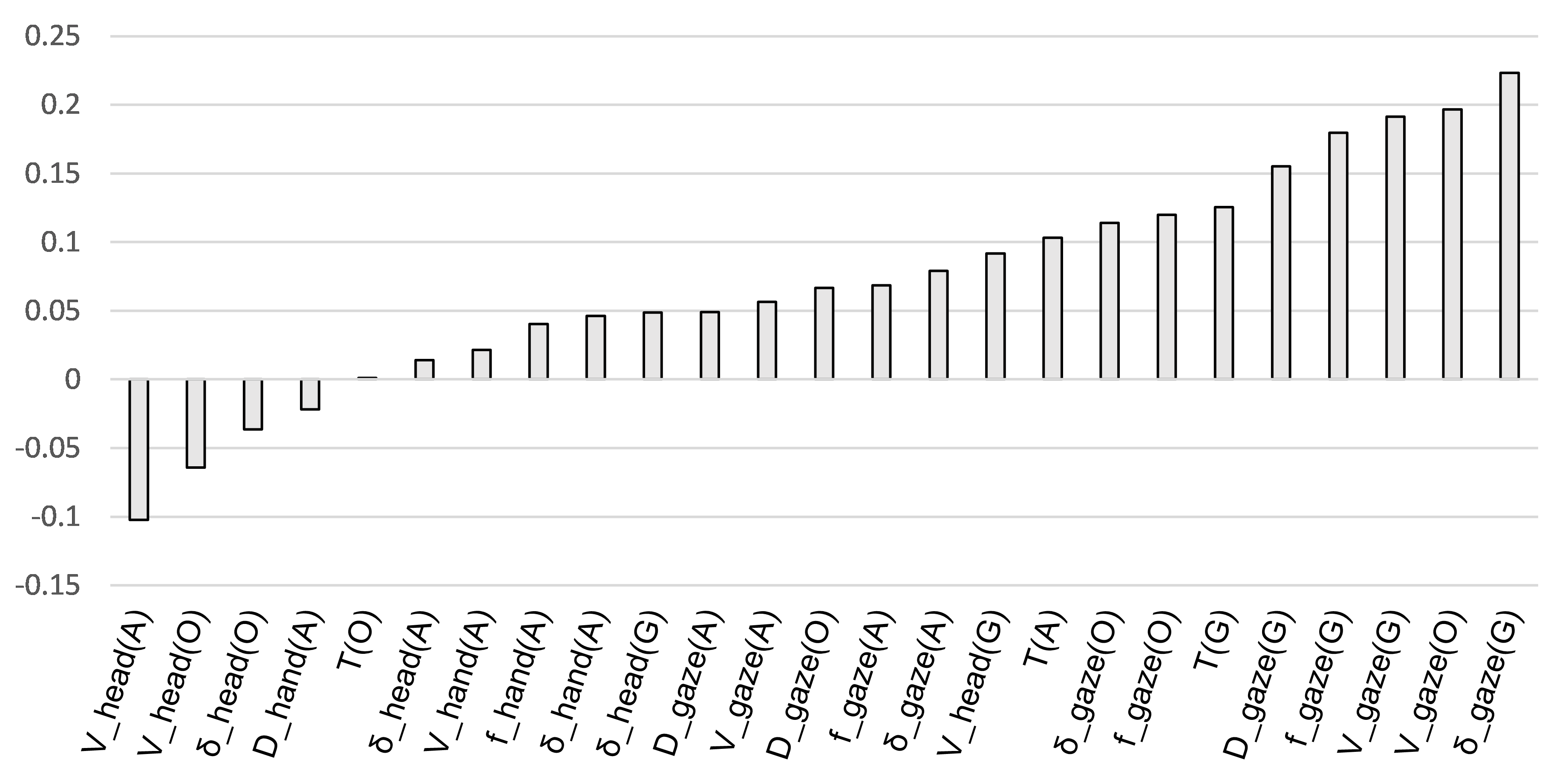}
\caption{Correlation of features to operation difficulties}
\label{fig:corrediff}
\end{figure}

Fig. \ref{fig:corrediff} depicts the correlation coefficients of the features and the user-rated difficulty scores over all the operation steps of both tasks.
The results show that \emph{gaze variance and velocity} of the G period are strongly correlated to operational difficulty.
This implies that the more difficult an operation step is, the more frequently the operator will search other regions prior to initiating operation.
In addition, faster gaze movement during O period will occur, which is probably due to result checking.
Conversely, if the user head moves faster during approaching and operation, the step was rated easier.
This is probably due to the operator's increasing familiarity and low concentration, as indicated by head movement.
Here, gaze and head in operation period showed opposite properties when indicating operational difficulty.

The above analysis provides clues for designing a metric to indicate operational difficulties for user guidance.
Although not all users require assistance with the difficult steps, providing support for most users on those difficult steps may enhance the task execution's overall efficiency.

\section{Task modeling}

\subsection{Prototype and Task Modeling}
As mentioned above, we aim to create adaptive guidelines to support users, and it is necessary to generate an extensive task model covering various operating methods and details.
To this end, we obtain learning resources from a sufficient amount of experiences performed by multiple users with different skill levels.
However, it is challenging to integrate various experiences directly into a single model because of the broad diversity of experience samples.
Different patterns of methods, orders, omissions, or repetitions can make a big difference between samples. 
Some experiences are even entirely incomparable with each other.

In \cite{chen2019intergrating}, we have solved this problem with a two-step approach.
First, build a baseline model from several manually-selected experts' operation experience.
Then, integrate experiences gradually with more considerable diversities into the baseline model.

The baseline model is crucial to the task modeling process because it provides a useful reference for aligning other experiences.
We first create the baseline mode as a template routine. Then the dynamic alignment approach is utilized to coordinate the sequence of operating procedures from the new-coming experiences.
In this way, a rich and diverse experience can be accurately integrated.
This method successfully modeled various task behaviors with online learning capability.
However, the experiences of the baseline model were manually selected.
It is vital to develop an automated method to select prototype experience from dozens or hundreds of experience records.
In this section, we describe a novel method for selecting high-skilled prototype experiences based on user behaviors.

\subsection{Prototype Selection Approach}
As mentioned above, features that are highly correlated to skills and less affected by interpersonal differences are regarded as reliable indicators of user skills.
Here, we propose a simple method to rank all experience skills by using several top-skill-correlated features.

Considering that user skills may not be evenly distributed across all experience procedures, we rank skills in each operation step (especially those difficult steps).
Besides, taking into account the data distribution among samples, such as the characteristics of \emph{accuracy} and \emph{representativeness} of the experiences, two global features are adopted to further improve performance.

The selection process for high-skilled experience is as follows:\\
(1) Each feature is used to obtain the ranks of all trial experiences of a task. If the feature is negative-correlated to the skill, the smaller the value, the higher the rank of the trial, and vice versa.\\
(2) Select several top-ranked trials corresponding to each feature (or each hotspot) to construct a pool of trials.\\
(3) The most frequently appeared trials in the pool are regarded as high-skilled prototype experiences.

In this process, we selected several top-ranked trials from each feature as the prototype candidates, and then use majority voting to find the final prototype among all candidates.
By doing so, the robustness of the results is enhanced because all the ultimately selected experiences are highly skilled that indicated by a combination of multiple features, which is believed to mitigate the influence of interpersonal differences.
The algorithm is described in Alg. \ref{code:bottomup}.

\begin{algorithm}[h]
\caption{Prototype chosen method}
\footnotesize
\begin{algorithmic}[1]
\REQUIRE ~~\\ 
1. The bag of hotspots of all interaction patterns appeared in task, $\bm{h} = \{h_1, h_2, \dots, h_M\}$.\\
2. The set of all experiences in a task, $\bm{e} = \{e_1, e_2, \dots, e_N\}$; any experience is a sequence of hotspots, ${e_j} = [h_{j1}, h_{j2}, \dots, h_{jm}]$.\\
3. The set of selected features ranked at K-top correlation to the user skill, $\bm{f} = \{f_1, f_2, \dots,f_K\}$.\\
\ENSURE ~~\\
A set of selected prototype experiences $\bm{e}_{opt}$.

\textbf{Begin}
\STATE Derive the ranks of a hotspot $h_i$ in all experiences by one select $f_k$, as $\gamma_{h_i,e_j}^{(f_k)} = ord(V_{h_i,e_j}^{(f_k)})$; where $V$ is the feature value, $ord$ is the order, and $e_j$ is the index of an experience contains this hotspot.
\STATE The average rank value of $h_i$ by all selected features is calculated as
$V_{h_i, e_j} = \alpha_{k}*\frac{1}{K} \sum_{k=1}^{K} \gamma_{h_i,e_j}^{(f_k)}$; where $\alpha_{k}$ is the assigned weight of $f_k$ according to its correlation score to skill.
\STATE Rank all the experiences with hotspot ${h_i}$, denote as $\gamma_{e_j}^{(h_i)} = ord(V_{h_i, e_j})$.
\STATE Construct a pool of experiences by the union of top-ranked experiences indicated by $\gamma_{e_j}^{(h_i)}$, as $\bm{p} =  [{\bm{e}}_{n}^{(h_1)} \cup {\bm{e}}_{n}^{(h_2)} \cup {\bm{e}}_{n}^{(h_3)} ,  \dots, {\bm{e}}_{n}^{(h_j)}]$; where $n$ is the number of selected experiences by a hotspot. \label{alg:pool}
\STATE The final ranks of experiences with all hotspots are indicated by their occurrences in the pool, as $\gamma_{e_j} = ord(\mathbb{F}_{e_j}^{(\bm{p})})$, where $\mathbb{F}$ is the number of occurrence. 
\STATE Select $q$ top-ranked experiences as the final prototype, $\bm{e}_{opt} = \bm{e}_{q}^{(\gamma_{e_j})}$.

\textbf{End}
\end{algorithmic}
\label{code:bottomup}
\end{algorithm}

\subsection{Global Features}
The above algorithm uses the combination of multiple low-level features to find the optimal prototype experiences. 
Each feature (or each hotspot) could serve as a weak indicator to identify high skills of experience, which is similar to Boosting \cite{boosting}.
In this section, we will also consider two global features to indicate high skills based on the global characteristics of the entire dataset, rather than local properties.

\textbf{1. Accuracy: } A highly skilled experience contains very little unnecessary interaction. \\
 An accurate prototype can ensure that the alignment process is successful for other experiences.
In general, the experience performed by highly skilled operators rarely employs unnecessary interactions, because operators tend to be economically (for example, with the fewest operating steps) and efficiently (for example, quickly operating on each step).

\textbf{2. Representativeness: } Typical high-skilled experiences have common interaction patterns, although the order of interactions may differ. \\
The prototype should represent the common method. If the prototype is too unique in the way of operation, then most of the operating experience will be difficult to unify.

We should note that the design of global features is based on the correctness of operations in all samples; that is, correct steps account for the majority of all steps performed by all users.
Even in the beginner's experience and their skill improvement process, it is considered easy to satisfy this condition.
Once most users can complete the task, the number of correct steps will basically exceed the number of incorrect steps.

The above two global features are calculated as follows.

\emph{a. The ratio of unnecessary interactions.}
We express the accuracy of the experience sequence by the ratio of unnecessary interactions.
Essential interactions are indispensable to task goals.
Almost every operator must perform these interactions to continue the task.
Unessential interaction is an unnecessary or harmful operation, such as a casual touch, a repetition, or an error.

\begin{figure}[htbp]
\centering
\includegraphics[width=0.45\textwidth]{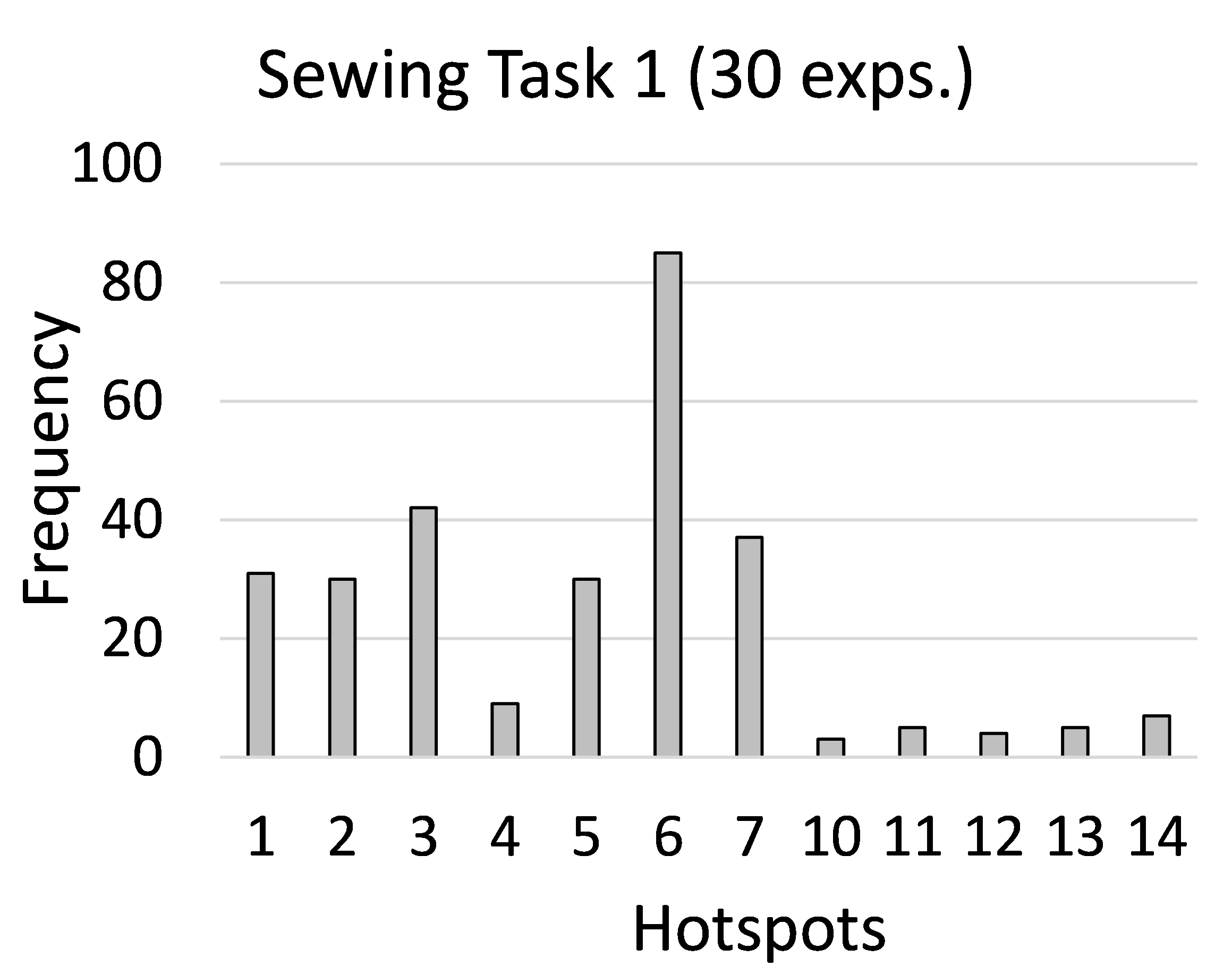} \quad \quad
\includegraphics[width=0.45\textwidth]{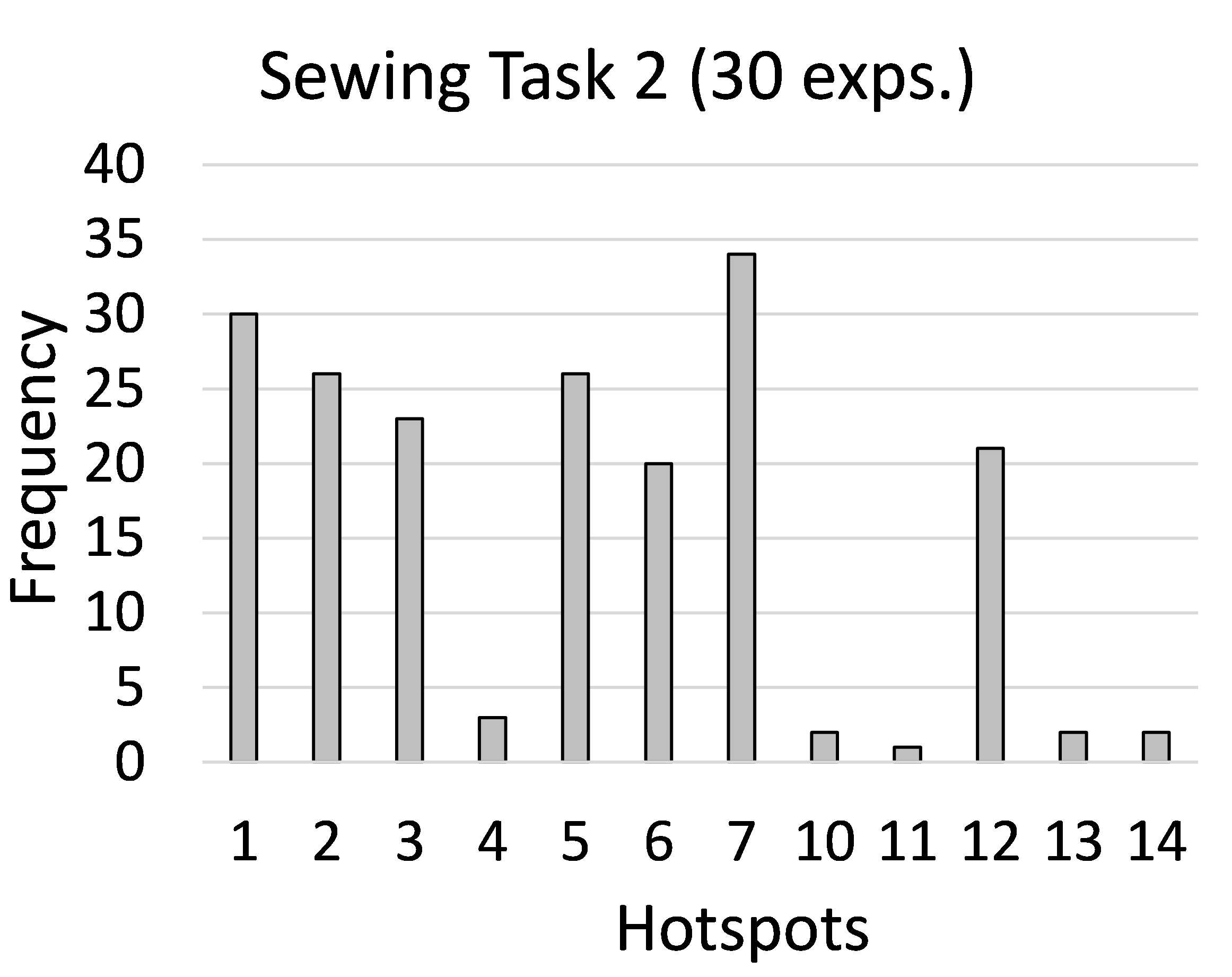}
\caption{Frequencies of operation patterns (hotspots) of 5 participants from two tasks.
The occurrence of hotspots \emph{1,2,3,5,6,7} and \emph{12} are much higher than half of the total experience number, while hotspots \emph{4,10,11,13} and \emph{14} are far below the half of total experience number.}
 \label{fig:hotoccur}
\end{figure}

Figure \ref {fig:hotoccur} shows histograms of the occurrence of hotspots of the two tasks. 
Note that the occurrence of hotspots is obvious in two categories: majority ($\geq50\%$) and minority ($\textless 50\%$).

These two categories have clear boundaries and can be used as clues to distinguish between essential and unessential interactions.
We can assert that essential interactions belong to the majority category.
Since almost every operator needs to perform the essential interaction, it should appear in most experiences.
On the other hand, the minority category contains the most unessential interactions; while it should not include any essential ones.

Based on this observation, we can consider the following steps to find an accurate experience:
First, distinguish the majority and minority hotspots by the interaction occurrences among all experiences.
Second, calculate the proportion of minority interactions in each experience.

\emph{b. Distance to the cluster center.}
Generally, a task requires a specific combination of operating modes and specific operating frequencies of these modes.
Most successful cases should have similar essential operating patterns; however, low-skilled experiences contain various errors and repetitions.

For example, most highly skilled users complete the task by operating hotspot A three times, hotspot B once, and hotspot C twice.
We consider \emph{(3, 1, 2)} as a standard operational attribute of these patterns.
Experience patterns and frequencies that are too far away from common attributes (such as \emph{(1, 4, 0)}) may be considered low-skilled or unique methods.

Based on this observation, we derive the representativeness of experience as follows:
First, convert all the experience sequences into bag-of-hotspots features.
Secondly, cluster all the new features, find the cluster center, and then calculate the Euclidean distance from each feature to the center.

The distance from the cluster center shows the representative score of an experience.
Meanwhile, the bag-of-hotspots feature represents each experience with equal length through the occurrences of hotspots, which only consider the combinations and frequencies of operation steps while removing the influence of different operational orders.

These two global features are added to enhance the prototype selection results in our experiment.

\subsection{Prototype Selection Results}
We selected the prototype experience from a small group of data: 60 experiences from 5 participants (30 experiences for each of the two sewing tasks).
In alg. \ref{code:bottomup} step \ref{alg:pool} to construct the experience pool, we choose 5 top-ranked experiences to put into the pool for each hotspot.
For the final prototype selection, we choose 3 top-ranked experiences.
In the global feature of experience accuracy, we set the occurrence threshold to 30\% of total experience number to distinguish between majority and minority interactions.
To evaluate the prototype selection results, we compare the selection results with the standard operating manual and calculate the F-score between the operating procedures.
According to the standard operating manual (ground truth), the DoF of task 1 is 2, with two order-changeable procedures. In contrast, task 2 involves two methods, each of which has two order-changeable procedures. The DoF is 4.

Table \ref{tab:prototype} shows the prototype detection results of the above algorithm.
Each method shows the selection results of the top-3 highly skilled experience.
The hotspot indexes represent the order of task steps.
In the table, ``AL'' means using all hotspots to rank experiences, ``DF'' means using only difficult hotspots to rank experiences.
``(3)'', ``(5)'' and ``(7)'' represent the number of top-skill-correlated features used to rank each hotspot.
``+GB'' represents adding global features.

\begin{table*}
\centering
\footnotesize
\vspace{-2.5cm}
\advance\leftskip-3cm
\advance\rightskip-3cm
\begin{threeparttable}
\caption{The detection of prototype experiences.}\label{tab:prototyperesult}
\begin{tabular}{|c|c|c|c|c|c|}
\hline
	 & Task 1 steps  & Task 2 steps    & \multirow{2}*{R} & \multirow{2}*{P} & \multirow{2}*{F} \\ 
	  &  (30 exp. Dof = 2) & (30 exp. Dof  = 4) & & & \\ \hline
	  
	  	& & & & & \\
	Ground Truth  & \multirow{2}*{\footnotesize{[1 2]  6  3  6 7 3  7  6  3  6}} & \footnotesize{[2 1] 5 7 12} & \multirow{2}*{1} & \multirow{2}*{1} & \multirow{2}*{1} \\ 
	(from manual) & \  & \footnotesize{5 7 12 [2 1]}  & \  & \  & \  \\ 
	& & & & & \\\hline
		& & & & & \\
{}         & \footnotesize{1     2       6     3     6      5     7     3       7     6     3       6     5}       & \footnotesize{7       3    12     1       2}             &     &     &     \\
{AL(3)}    & \footnotesize{2     4     5      13    13     6       6     6     3      13     6     6       5     1} & \footnotesize{5     7       3    12    12       1     4} & {0.842} & {0.742} & {0.789} \\
{}         & \footnotesize{2     1     5       7     6     3       6     7     3       7     3     6}               & \footnotesize{5     7       3    12     1       2}       &     &     &     \\
{}         &                                                                                       &                                         &     &     &     \\
{}         & \footnotesize{1     2     6       3     6     5       7     3     7       6     3     6       5}       & \footnotesize{7     3    12       1     2}               &     &     &     \\
{AL+GB(3)} & \footnotesize{1     2       6     3     6       5     7     3       7     6     3       6     5}       & \footnotesize{7    12       1     2}                     & {0.791} & {0.745} & {0.767} \\
{}         & \footnotesize{2     4       5    13    13       6     6     6       3    13     6       6     5     1} & \footnotesize{1     2       7     3}                     &     &     &     \\
{}         &                                                                                       &                                         &     &     &     \\
{}         & \footnotesize{2     1     5       7     6     3       6     7     3       7     3     6}               & \footnotesize{5     7       3    12    12       1     4} &     &     &     \\
{DF(3)}    & \footnotesize{1     2       6     3     6       5     7     3       7     6     3       6     5}       & \footnotesize{7     3    12       1     2}               & {0.903} & {0.810} & {0.854} \\
{}         & \footnotesize{2     1       5     6     3       6     7     7       5     6     3       6}             & \footnotesize{5     7       3    12     1       2}       &     &     &     \\
{}         &                                                                                       &                                         &     &     &     \\
{}         & \footnotesize{1     2       6     3     6       5     7     3       7     6     3       6     5}       & \footnotesize{7     3      12     1     2}               &     &     &     \\
{DF+GB(3)} & \footnotesize{2     1       5     7     6       3     6     7       3     7     3       6}             & \footnotesize{5     7       3    12    12       1     4} & {0.885} & {0.807} & {0.844} \\
{}         & \footnotesize{1     2       6     3     6       5     7     3       7     6     3       6     5}       & \footnotesize{5      7      2        1}                  &     &     &     \\
{}         &                                                                                       &                                         &     &     &     \\\hline
	& & & & & \\
{}         & \footnotesize{1     2     6       3     6     5       7     3     7       6     3     6       5}       & \footnotesize{5     7       3    12     1       2}       &     &     &     \\
{AL(5)}    & \footnotesize{2     4     5      13    13     6       6     6     3      13     6     6       5     1} & \footnotesize{7     3      12     1     2}               & {0.809} & {0.736} & {0.771} \\
{}         & \footnotesize{1     2       5     6     3       6     7     7       6     3     6}                     & \footnotesize{5     7       3     1     4}               &     &     &     \\
{}         &                                                                                       &                                         &     &     &     \\
{}         & \footnotesize{1     2     6       3     6     5       7     3     7       6     3     6       5}       & \footnotesize{7     3      12     1     2}               &     &     &     \\
{AL+GB(5)} & \footnotesize{2     4       5    13    13       6     6     6       3    13     6       6     5     1} & \footnotesize{5     7       3    12     1       2}       & {0.842} & {0.769} & {0.804} \\
{}         & \footnotesize{1     2       5     6     3       6     7     7       6     3     6}                     & \footnotesize{7     3      12     1     2}               &     &     &     \\
{}         &                                                                                       &                                         &     &     &     \\
{}         & \footnotesize{1     2     6       3     6     5       7     3     7       6     3     6       5}       & \footnotesize{2     1       5    12     7      12}       &     &     &     \\
{DF(5)}    & \footnotesize{2     1     5       6     3     6       7     7     5       6     3     6}               & \footnotesize{7     3      12     1     2}               & \textbf{0.952} & \textbf{0.844} & \textbf{0.894} \\
{}         & \footnotesize{2     1       5     6     3       6     7     3       7     6     3       6}             & \footnotesize{5     7       3    12     1       2}       &     &     &     \\
{}         &                                                                                       &                                         &     &     &     \\
{}         & \footnotesize{1     2     6       3     6     5       7     3     7       6     3     6       5}       & \footnotesize{7     3      12     1     2}               &     &     &     \\
{DF+GB(5)} & \footnotesize{1     2       6     3     6       5     7     3       7     6     3       6     5}       & \footnotesize{2     1       5    12     7      12}       & \textbf{0.933} & \textbf{0.829} & \textbf{0.878} \\
{}         & \footnotesize{1     2     6       3     6     5       7     3     7       6     3     6       5}       & \footnotesize{5     7       2     1}                     &     &     &     \\
{}         &                                                                                       &                                         &     &     &     \\\hline
	& & & & & \\
{}         & \footnotesize{2     1       5     7     6       3     6     7       3     7     3       6}             & \footnotesize{5     7       3    12     1       2}       &     &     &     \\
{AL(7)}    & \footnotesize{1     2       5     6     3       6     7     7       6     3     6}                     & \footnotesize{5     7       3     1     4}               & {0.855} & {0.814} & {0.834} \\
{}         & \footnotesize{1     2       5     6     3       6     7     7       6     3     6}                     & \footnotesize{7     3      12     1     2}               &     &     &     \\
{}         &                                                                                       &                                         &     &     &     \\
{}         & \footnotesize{1     2       6     3     6       5     7     3       7     6     3       6     5}       & \footnotesize{7     3      12     1     2}               &     &     &     \\
{AL+GB(7)} & \footnotesize{2     1       5     7     6       3     6     7       3     7     3       6}             & \footnotesize{5     7       3    12     1       2}       & {0.870} & {0.804} & {0.835} \\
{}         & \footnotesize{1     2       5     6     3       6     7     7       6     3     6}                     & \footnotesize{5     7       3     1     4}               &     &     &     \\
{}         &                                                                                       &                                         &     &     &     \\
{}         & \footnotesize{1     2       6     3     6       5     7     3       7     6     3       6     5}       & \footnotesize{7     3      12     1     2}               &     &     &     \\
{DF(7)}    & \footnotesize{2     1     5       6     3     6       7     7     5       6     3     6}               & \footnotesize{5     7       3    12     1       2}       & \textbf{0.952} & \textbf{0.844} & \textbf{0.894} \\
{}         & \footnotesize{2     1       5     6     3       6     7     3       7     6     3       6}             & \footnotesize{5     7       3    12     1       2}       &     &     &     \\
{}         &                                                                                       &                                         &     &     &     \\
{}         & \footnotesize{1     2       6     3     6       5     7     3       7     6     3       6     5}       & \footnotesize{7     3      12     1     2}               &     &     &     \\
{DF+GB(7)} & \footnotesize{2     1     5       6     3     6       7     7     5       6     3     6}               & \footnotesize{5     7       2     1}                     & {0.885} & \textbf{0.833} & \textbf{0.858} \\
{}         & \footnotesize{2     1       5     6     3       6     7     3       7     6     3       6}             & \footnotesize{7    12       1     2}                     &     &     &    \\
	& & & & & \\
\hline
\end{tabular}
\label{tab:prototype}
\begin{tablenotes}
   \item \tiny{``[ ]'' contains order-changeable steps. Step indexes: 1. \emph{speed button}, 2. \emph{sewing pattern}, 3. \emph{operate on cloth}, 4. \emph{thread setting button}, 5. \emph{needle position button}, 6. \emph{start/stop button}, 7. \emph{lever}, 10. \emph{top side}, 11. \emph{left side}, 12. \emph{cut thread}, 13. \emph{reverse stitch button}, 14. \emph{threading pad}. ``AL'' represents all hotspots, ``DF'' represents only difficult hotspots.
``(3)'', ``(5)'' and ``(7)'' represent the number of top-skill correlated features.
``+GB'' represents global features.}
        \end{tablenotes}
     \end{threeparttable}
\end{table*}

Compared with the standard manual, the average recall rates of selected prototypes of all methods of task 1 and task 2 are 0.91 and 0.88, respectively, and the precision of task 1 and task 2 are 0.80 and 0.81, respectively.
Considering the number of skills-related features for selecting the final prototype, combining the top-5 features can get the best prototype selection results.
Due to the influence of two additional features, the result is slightly degraded (top-7).

Taking into account the use of different types of hotspots, the method using only difficult hotspots (DF) achieves the best F-score regardless of how many skill-related features are used.
Adding the global features (+GB) does not significantly improve the DF method; however, it enhances the performance of using all hotspots (AL).

This result shows that difficult steps are a better indicator of skill level than just using all steps, which is sufficient to select a good prototype experience.
In addition, global features can be used as a good indicator of skills when lacking the information about step difficulties.

We can see that the final selected prototype experiences with the DF method are very close to the standard procedure with all the essential procedures recalled.
However, none of the participants' actual operating experience is exactly the same as the manual.

\subsection{Task Modeling Results}
\begin{figure*}[htbp]
\centering
\advance\leftskip-2cm
\advance\rightskip-2cm
\vspace{-2.5cm} 
\begin{minipage}[b]{0.7\textwidth}
\footnotesize
\makebox[\textwidth][c]{\subfloat[Groundtruth(1)]{\includegraphics[width=1\textwidth]{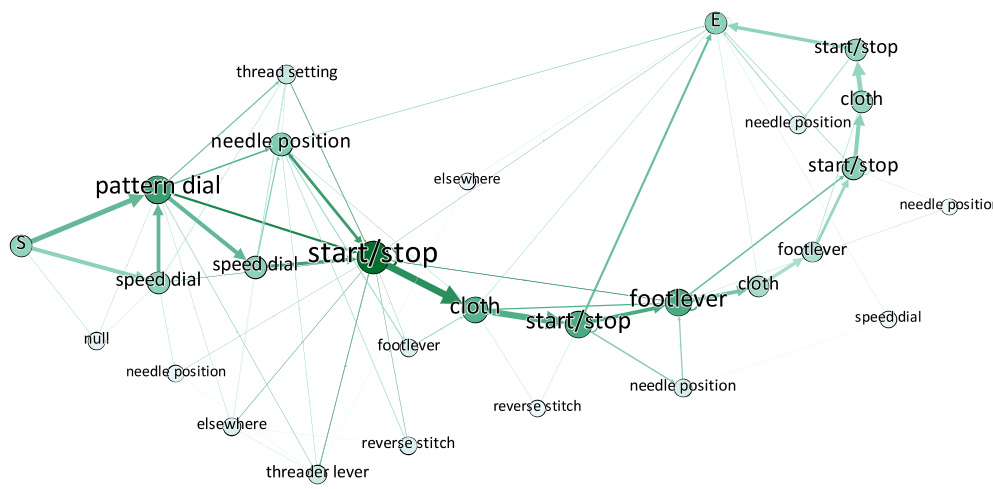}}}\\
\makebox[\textwidth][c]{\subfloat[Detected(1)]{ \includegraphics[width=1\textwidth]{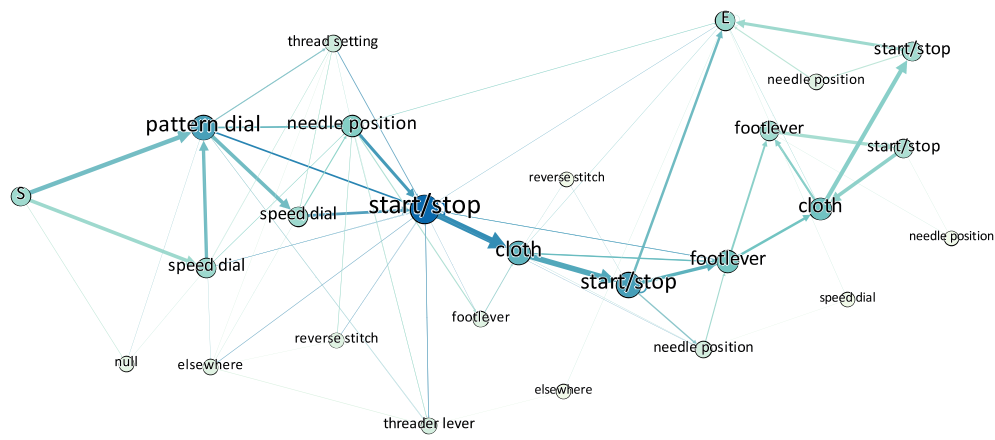}}}\\
\makebox[\textwidth][c]{\subfloat[Random(1)]{ \includegraphics[width=0.92\textwidth]{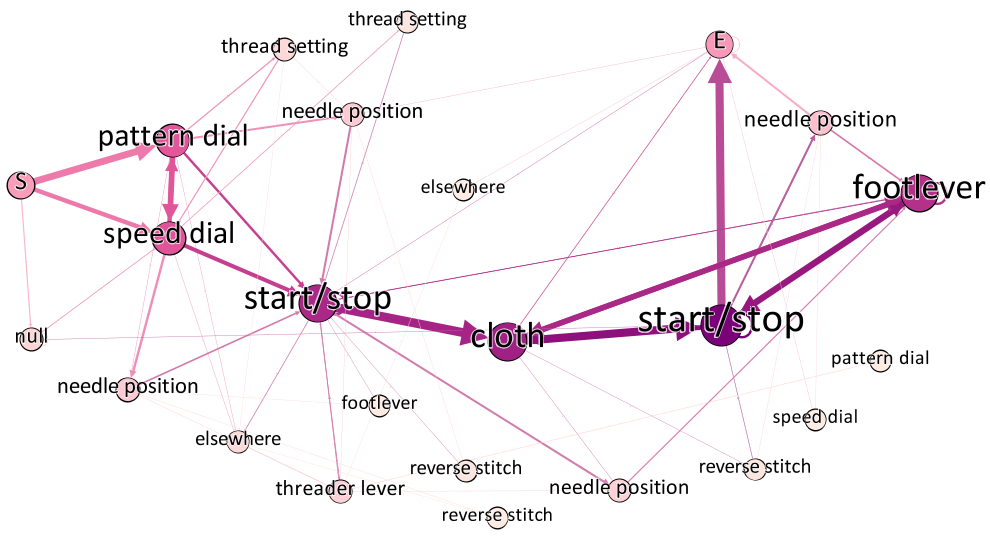}}}\\
\end{minipage}
\ \ \ \ \ \ \ \ \ \ \ \ 
\begin{minipage}[b]{0.39\textwidth}
\footnotesize
\makebox[\textwidth][c]{\subfloat[Groundtruth(2)]{\includegraphics[width=0.92\textwidth]{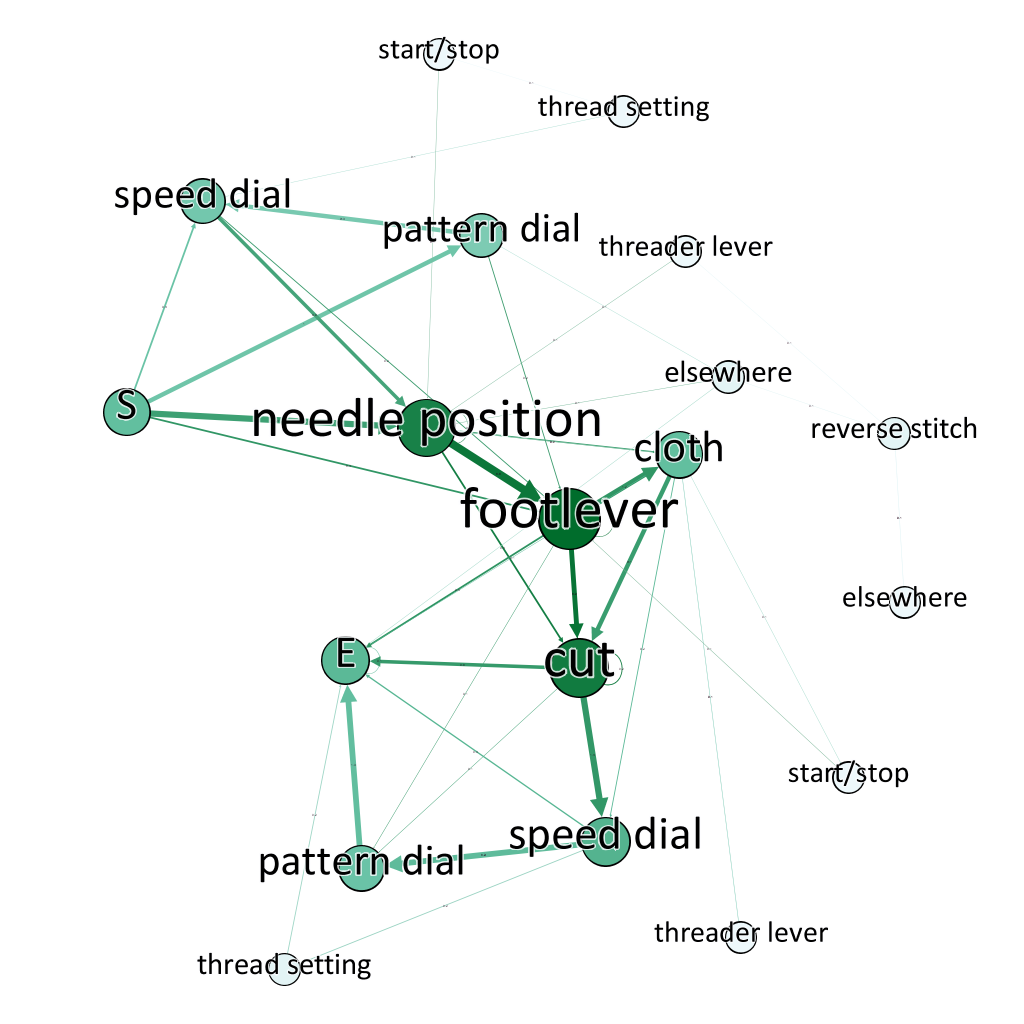}}}\\
\makebox[\textwidth][c]{\subfloat[Detected(2)]{ \includegraphics[width=0.92\textwidth]{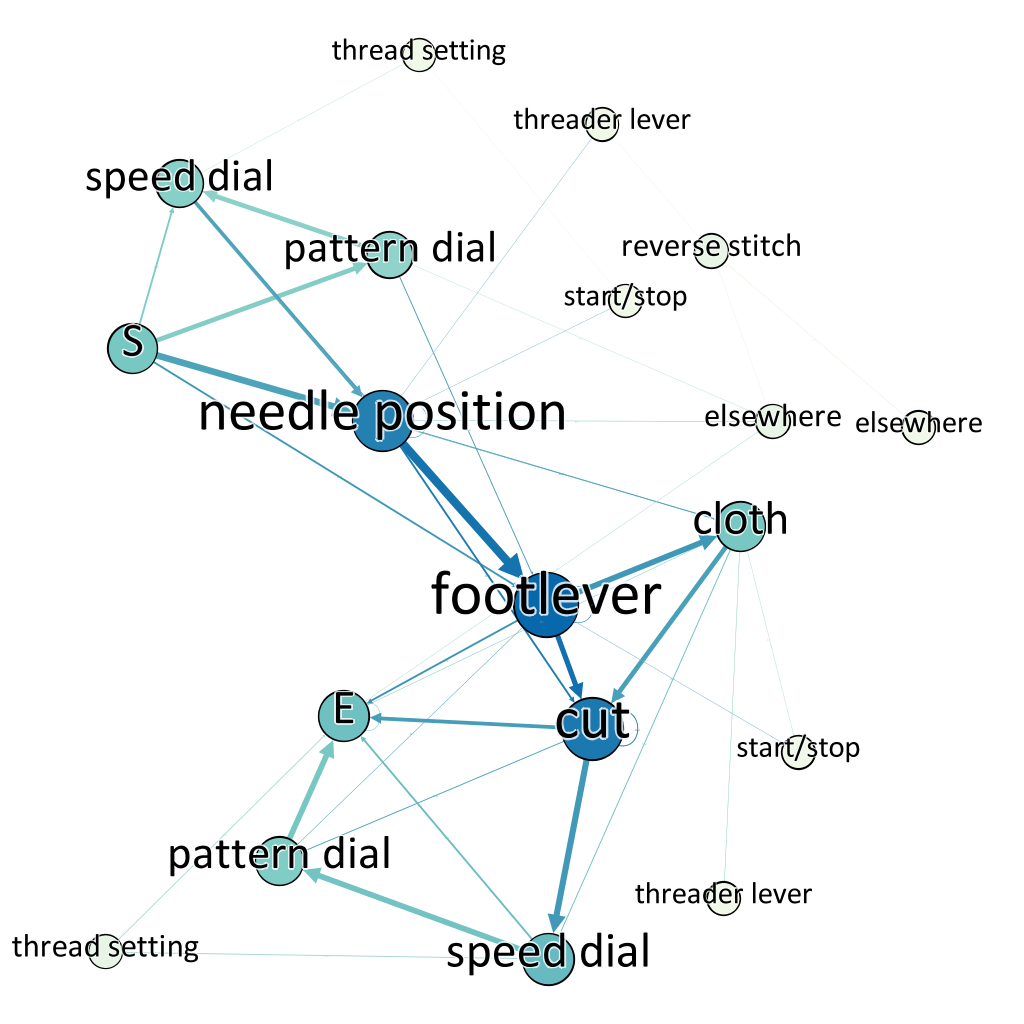}}}\\
\makebox[\textwidth][c]{\subfloat[Random(2)]{ \includegraphics[width=0.92\textwidth]{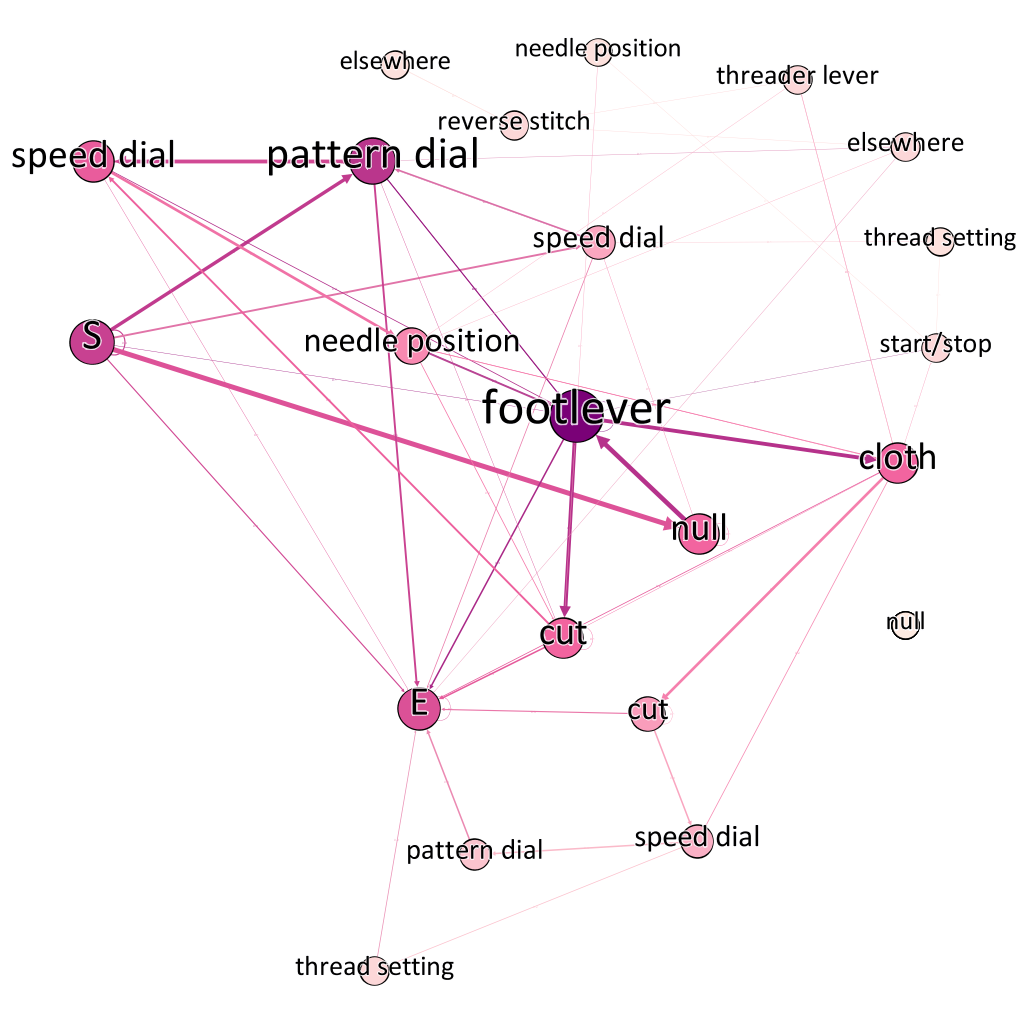}}}
\end{minipage}
\caption{The models for integrating experiences for two sewing tasks with different prototypes (30 experiences for each task).
From top to bottom are unified models with manually-selected baseline (a, d), automatically selected experience with the DF method as baseline (b, e), and the prototype by randomly choosing an experience as the baseline (c, f).
The saturation of the nodes indicates the sum of the In-Out transition probabilities of the nodes in HMM.}
\label{fig:models}
\end{figure*}

We can use the selection results to build a baseline model for experience integration and task modeling.

Figure \ref{fig:models} illustrates the task model built using different prototypes for two tasks.
We compared task models built with (a) standard manual prototypes (ground truth), (b) selected high-skilled prototypes (by our algorithm), and (c) randomly-selected prototypes.

The task models built using our prototype selection approach are very similar to the ground-truth model in both states and transitions.
While the randomly-selected prototype models lack some essential states, the operation methods are incomplete, and some experience alignment is unsuccessful.
When an essential operation is missing in the alignment forward routine. In that case, the current mode may jump backward to find a state with the same observation or fail to see any corresponding state.

We can conclude that our prototype selection method takes into account the properties of the operations, that is, skill relevance, operation difficulty, and global attributes, can automatically find a good baseline experience for task modeling.

\section{Conclusion}
In this paper, we proposed an approach to model the multiformity of the task and analyze users' behavior at a variety of skill levels by considering the properties of user--machine interactions and interpersonal differences.
We systematically acquired relevant samples and, on this basis, analyzed the characteristics of operational behavior patterns with the features extracted from the information about user gaze, head movements, hand movements, and hotspot interactions concerning both temporal and spatial domains.
The analysis of user behaviors is used to select prototype baseline experiences for task modeling.
The experimental result shows some features are useful indexes of operator skill levels and operational difficulties, particularly for task duration, head movement, and gaze properties.
We also demonstrated that an unsupervised prototype selection approach could be adopted to derive an extensive task model for guidance.
In future, we need to design metrics based on user skill levels, operational difficulties, and interpersonal differences for designing adaptive user assistant systems.

\end{document}